\documentclass[12pt]{article} 

\newenvironment{pf}{\noindent{\bf Proof:}}{\newline\kbn}

\usepackage{amssymb,mathrsfs}
\usepackage{german,a4,isolatin1}
\newtheorem{satz}{Theorem}
\newtheorem{folge}[satz]{Corollary}
\newtheorem{lemma}[satz]{Lemma}
\newtheorem{prop}[satz]{Proposition}
\newtheorem{defi}[satz]{Definition}
\newtheorem{ass}{Additional Assumption}

\newcommand{\Einsop}{\leavevmode{\rm 1\mkern  -4.4mu l}}
\newcommand{\Seins}{\mathsf{S}^1}

\newcommand{\Diff}{{\mbox{\it Diff}}}

\newcommand{\PSL}{{\mbox{\it PSL}}}
\newcommand{\lie}[1]{{\mathfrak #1}}
\newcommand{\komm}[2]{{\left[ #1 , #2 \right]}} 
\newcommand{\betrag}[1]{{\left| #1 \right|}} 
 
\newcommand{\klammer}[1]{{\left( #1 \right)}}
\newcommand{\skalar}[2]{{\left\langle #1 , #2 \right\rangle }}
\newcommand{\lok}[1]{{\mathcal #1}}
\newcommand{\Hilb}[1]{{\mathcal #1}}
\newcommand{\Name}[1]{{\sc #1}} 
\newcommand{\dopp}[1]{{\mathbb #1}}
\newcommand{\symb}[1]{{\mathbf #1}}
\newcommand{\opo}{{1\!+\!1}}

\newcommand{\kbn}{$\square$}

\begin{document}

\title{Local Nature of \Name{Coset} Models}
\author{S\o{}ren K"oster\\Inst.\ f.\ Theor.\ Physik\\ Universit\"at
  G\"ottingen\\ Tammannstr. 1, 37077 G\"ottingen \\ Germany}
\date{} 

\maketitle
\begin{abstract} 
The local algebras of the maximal \Name{Coset} model $\lok{C}_{max}$
associated with a 
chiral conformal subtheory $\lok{A}\subset\lok{B}$ are shown to
coincide with the local relative commutants of $\lok{A}$ in $\lok{B}$,
provided $\lok{A}$ possesses a stress-energy tensor.

Making the same assumption, the adjoint action of the unique
inner-implementing representation 
$U^\lok{A}$ associated with $\lok{A}\subset\lok{B}$ on  the local
observables in $\lok{B}$ is found to define net-endomorphisms of
$\lok{B}$. This property is exploited for constructing from $\lok{B}$
a conformally covariant holographic image in $1+1$ dimensions which
proves useful as a geometric picture for the joint inclusion
$\lok{A}\vee\lok{C}_{max}\subset\lok{B}$.  

Immediate applications to the analysis of current subalgebras are
given and the relation to normal canonical tensor product subfactors
is clarified. A natural converse of \Name{Borchers}' theorem on
half-sided translations is made accessible.\\[1mm]
\end{abstract}

\section{Introduction}
\label{sec:intro}

Structural and conceptual questions of quantum field theory are often
addressed best within the framework of local quantum physics,  where
physics is described by assigning local algebras $\lok{B}(\lok{O})$ of
observables to localisation regions $\lok{O}$ rather than in terms of
quantum fields \cite{rH92}. In this picture it is natural to
investigate the relative position of subnets
$\lok{A}(\lok{O})\subset\lok{B}(\lok{O})$ in the larger theory
$\lok{B}$. 

In $3+1$-dimensional spacetime this problem may be dealt with by means
of the powerful reconstruction method of \Name{Doplicher and Roberts}
\cite{DR90} and according to the results of \Name{Carpi and Conti}
\cite{CC01} any subtheory satisfying certain assumptions
is in a tensor product position in the larger theory. In lower
dimensions, however, the situation is less restrictive and a lot of
interesting examples are known.

In this article we study a class of inclusions of local quantum
theories $\lok{A}\subset \lok{B}$, where $\lok{B}$ is given in its
vacuum representation. When not all of the energy-content of $\lok{B}$
belongs to $\lok{A}$, there is space for other subtheories
$\lok{C}\subset\lok{B}$ which commute with all of $\lok{A}$. We call
such subtheories $\lok{C}$ {\em {\em \Name{Coset} models (associated with
  $\lok{A}\subset\lok{B}$)}}, and we  want to derive typical features
of these. By placing the problem into the setting 
of chiral conformal quantum field theory the large spacetime symmetry
enables us to state and discuss the problems concerned clearly and
rigorously.

Chiral conformal \Name{Coset} models are 
studied for various reasons, in most cases connected to inclusions of  chiral
current algebras. These models exhibit a rich and yet tractable
structure. One of the major achievements in this direction was the
construction of the discrete series of \Name{Virasoro} theories as
\Name{Coset} models by \Name{Goddard, Kent, and Olive}
\cite{GKO86}. In the algebraic approach
to quantum field theory much has been achieved for inclusions of local
quantum theories generated by chiral current algebras and closely
related structures 
\cite{tL94,fX00,fX99,fX01, rL01,KL02}. 

Here, we want to broaden the perspective by using methods which do not
make use of structures specifically connected to chiral current
algebras, but which apply in a more general context: For chiral
conformal quantum field theories the natural localisation 
regions are open, non-dense intervals $I$ in the  circle.
The  local
algebras $\lok{C}(I)$ of a \Name{Coset} model  associated with a subnet
$\lok{A}\subset\lok{B}$ are contained in the local relative 
commutants  $\lok{C}_I := \lok{A}(I)'\cap \lok{B}(I)$. Actually, if
the local relative commutants fulfill isotony, ie the local relative
commutants  
$\lok{C}_I$ increase with $I$, the $\lok{C}_I$
 define a \Name{Coset} model themselves which is obviously maximal. 

Isotony for the  $\lok{C}_I$ holds for chiral current subalgebras
because of the strong 
additivity property of these models \cite{vL97} (corollary IV.1.3.3.): If $I_1$,
$I_2$ arise from $I$ by removing a point in its interior,  the local algebra
$\lok{A}(I)$ is generated by 
its subalgebras $\lok{A}(I_1)$ and $\lok{A}(I_2)$. This property is
absent in many chiral conformal models \cite{BS90,jY94} and,
naturally, the question arises, under which circumstances the
equality of $\lok{C}_I$ and $\lok{C}_{max}(I)$ can be proven. Our
intention is to find more general conditions which secure this equality for
various reasons and we refer to this task as the {\em isotony problem}. 

In a recent work \cite{sK02} we
constructed for any chiral conformal subtheory $\lok{A}\subset\lok{B}$
a globally $\lok{A}$-inner representation $U^\lok{A}$ which implements
the (global)
chiral conformal transformations on $\lok{A}$. The result
provides a factorisation of $U$, the implementation of chiral
conformal symmetry in the vacuum representation of $\lok{B}$,  into
two commuting representations, 
$U^\lok{A}$ and $U^{\lok{A}'}$, which share with $U$ the properties of
leaving the vacuum invariant and of positivity of energy. It was
proved, by a simple argument, that the local operators in $\lok{B}$
which commute with $U^\lok{A}$ form the maximal \Name{Coset} model
$\lok{C}_{max}$ 
associated with every particular inclusion $\lok{A}\subset\lok{B}$. 
 $\lok{C}_{max}$ can be non-trivial, if we have
$U^{\lok{A}'}\neq \Einsop$, ie if not all of the energy-content of $\lok{B}$
belongs to $\lok{A}$.

We would like to have a simple and applicable characterisation
of local operators in $\lok{B}$ which belong to a \Name{Coset} model
associated with a subtheory $\lok{A}$, and we want this
characterisation to involve only {\em local} data according to the
conviction that all observation is of finite extension and of finite
duration. Of course, it is in principle possible to make this decision
 simply by taking all operators from $\lok{C}_I$ and discarding all
 operators which do not commute with all operators belonging to an
 algebra $\lok{A}(J)$, $J$ slightly enlarged. But chiral conformal
 quantum field theories usually behave well when $J$ tends to $I$, and
 we are led to the conjecture that the local algebras of the maximal
 \Name{Coset} model and the corresponding local relative commutants
 should coincide in very general circumstances.
As it stands for the moment, the maximal \Name{Coset} model is
determined by global data, the inner-implementing representation
$U^\lok{A}$, and establishing the equality
$\lok{C}_I=\lok{C}_{max}(I)$ would prove that the \Name{Coset} model
is of a {\em local nature}, its local operators being singled out by a
simple algebraic relation only involving local data associated with
the very same localisation region. 

For dealing with the isotony problem in our context\footnote{Apparently,
  \Name{Carpi and Conti} encountered the same problem while
  generalising their analysis \cite{CC01} to general field algebras
  and solved it by methods quite different from the ones applied here
  \cite{CC03}.}, we look at the
action of   $Ad_{U^\lok{A}}$ on the local observables of $\lok{B}$. 
Because the  
 construction of $U^\lok{A}$ does not refer to the local structure  of
 $\lok{A}$ at all, we need some information on the way this
 representation is generated by local observables. In chiral conformal
 field theory it is natural to assume that the inner implementing
 representation is generated by integrals of a stress-energy tensor
 affiliated with $\lok{A}$. This assumption does not imply strong
 additivity \cite{BS90} and concerning the models known today (at
 least to the author) is more general, since all strongly additive
 models contain a stress-energy tensor. 

Because of the special features stress-energy tensors of chiral
(and $\opo$-dimensional) conformal field theory have according
to the
\Name{L"uscher-Mack} theorem \cite{FST89}, solving the isotony problem
proves possible, but the presence of a stress-energy tensor does not 
trivialise it at all. In fact, one is led to pinpoint the
problem very much using arguments independent of the additional
assumption, before the
stress-energy tensor actually is needed to prove two crucial, but
natural lemmas. Our discussion should, therefore, serve well as a
setup for further generalisations.

Even for current subalgebras, which always contain a stress-energy tensor
by the \Name{Sugawara} construction, the action of a stress-energy
tensor $\Theta^\lok{A}$ of a current subalgebra on general currents
in the larger current algebra $\lok{B}$ has not been studied as such,
yet. Only in connection with the classification of {\em conformal
inclusions}, ie the case that the stress-energy tensor
$\Theta^\lok{B}$ coincides with that of $\lok{A}$
\cite{SW86,AGO87,BB87}, this action has
been object of research. The new
perspective of analysing the action of $U^\lok{A}$ on $\lok{B}$ (in
this context: of $\Theta^\lok{A}$ on $\lok{B}$) 
directly has led to a simple and natural characterisation of
conformal inclusions by methods familiar in (axiomatic) quantum field
theory \cite{sK03a}. 

As mentioned above, the {\em local nature} of maximal \Name{Coset}
models associated with current subalgebras $\lok{A}\subset\lok{B}$ is
clear because of the strong additivity property of $\lok{A}$. If the
embedding $\lok{A}\vee\lok{C}_{max}\subset\lok{B}$ is known to be of
finite index , then $\lok{C}_{max}$ inherits the strong additivity
property from $\lok{A}$ and $\lok{B}$ by results of \Name{Longo}
\cite{rL01}.  According to \Name{Xu} \cite{fX00} a large number of
current algebra inclusions are known to 
satisfy this condition ({\em cofinite} inclusions
$\lok{A}\subset\lok{B}$), but for a lot of others the situation has not 
been clarified, yet. If we now look at the embedding
$\lok{C}_{max}\subset\lok{B}$ and consider the (``iterated'')
\Name{Coset} models associated with this inclusion, we arrive at
another isotony problem. In case 
$\lok{C}_{max}$ is strongly additive as well, the local relative
commutants of $\lok{C}_{max}$  and the local algebras of
$\lok{C}_{max}$ form a pair of subnets which locally are their
respective relative commutants. Inclusions of this type are of particular
interest and \Name{Rehren} called them a {\em normal} pair of subnets
\cite{khR00}. 

Our analysis applies to the maximal \Name{Coset} models associated with
current subalgebras as these always
contain the \Name{Coset} stress-energy tensor
$\Theta^\lok{B}-\Theta^\lok{A}$. This way we extend the finding on
normal pairs for cofinite current subalgebras to all inclusions  
$\lok{A}\subset\lok{B}$ where both $\lok{B}$ 
and $\lok{A}$ contain a stress-energy tensor, independent
of strong additivity or the index of the inclusion
$\lok{A}\vee\lok{C}_{max}\subset\lok{B}$. 

In the next section we first state our general assumptions and
conventions and then discuss the ``geometric impact'' of
$U^\lok{A}$ on $\lok{B}$. Intuitively, we do not expect an observable
of $\lok{B}$ to be more sensitive to the action of $Ad_{U^\lok{A}}$
than to that of $Ad_{U}$: the generator of translations, $P$, is known
to decompose into two commuting positive parts,
$P=P^\lok{A}+P^{\lok{A}'}$, and regarding them as chiral analogues of
\Name{Hamilton}ians leads us to the expectation that $P^\lok{A}$
should not transport observables of $\lok{B}$ ``faster'' than $P$
itself. A typical local observable $B$ in $\lok{B}$ should exhibit a
behaviour interpolating between invariance ($B$ in $\lok{C}_{max}$)
and covariance ($B$ in $\lok{A}_{max}$). 

For this behaviour to be ensured we have, as it turns out, only to
show that scale transformations represented through $U^\lok{A}$
respect the two fixed points of scale transformations, namely $0$ and
$\infty$, when acting on $\lok{B}$. We can prove this to be the case
in presence of a stress-energy tensor and it seems natural in any
case. The sub-geometrical
transformation behaviour for translations, which we expect, then
follows by results of \Name{Borchers} \cite{hjB97,hjB97a} using the
spectrum condition and modular theory. We collected, rearranged and
reformulated 
results of \Name{Borchers} and \Name{Wiesbrock} in order to provide a
natural {\em converse of {Borchers}' theorem on half-sided
translations}, which was not yet available in the literature. By
extending the 
analysis to general conformal transformations we arrive at the notion
of {\em net-endomorphism property} for the action of $U^\lok{A}$ on
$\lok{B}$.

In the third section we use the net-endomorphic action of $U^\lok{A}$
to construct from the chiral conformal theory $\lok{B}$ a conformal
net in $\opo$ dimensions which contains the chiral algebras as
{\em time-zero algebras}. The result satisfies all axioms of a
$\opo$-dimensional conformal quantum theory, except that its
translations in spacelike directions to the right have positive
spectrum rather than in 
future-like directions. While this prohibits interpreting the  picture
of {\em chiral holography} 
as (completely) physically sensible, it gives a satisfactory geometric
interpretation to the net-endomorphisms induced by  $U^\lok{A}$ and it
provides a rather helpful 
geometrical framework of a {\em quasi-theory} in $\opo$ dimensions. The
subnet $\lok{A}\subset\lok{B}$ and its \Name{Coset} 
models appear as subtheories of {\em chiral observables}
and thus we make connection with results of \Name{Rehren}
\cite{khR00}, which have interesting consequences for known examples.

In the closing section we provide our solution to the isotony problem
(main theorem \ref{th:main}),
ie we establish the local nature of the maximal \Name{Coset}
model. We start by giving a new characterisation of $\lok{C}_{max}$
making use of the particular structure of the group of chiral
conformal transformations. And then, again, the presence of a
stress-energy tensor for $\lok{A}$ is  
only needed in order to establish a rather natural, but crucial lemma
on the representation of scale transformations through
$U^\lok{A}$.   At the very end we discuss possible generalisations to
models having no stress-energy tensor and to subtheories in other
spacetimes. The appendix contains background on our additional assumption on
the inner-implementing representation $U^\lok{A}$, while  we will use
an abstract formulation of it in the main sections, and a simple,
technical lemma on scale transformations as elements of the group of
orientation preserving diffeomorphism on the circle,
$\Diff_+(\Seins)$.

\section{Net-endomorphism property}
\label{sec:netend}

The fundamental object of this study is an inclusion of a chiral
conformal theory, $\lok{A}$, in another chiral
conformal theory,  $\lok{B}$. The theory $\lok{B}$ shall be given
in its vacuum representation, of which we summarise the general
assumptions and some of its properties (cf \cite{GL96,FG93} and
references therein), and we describe the embedding of $\lok{A}$ in
$\lok{B}$ in this setting.

The localisation regions for chiral conformal theories are taken to be
the {\em proper intervals} contained in the unit circle $\Seins$,
which is to be regarded as the conformal compactification of a (chiral)
light-ray; the point $+1$ on $\Seins$ corresponds to the point $0$ on
the light-ray and $-1\in \Seins$ corresponds to $\infty$. A proper interval $I$ is an open, connected subset of
$\Seins$ which has a {\em causal (open) complement}, $I':= \{\Seins\setminus
I\}^\circ\neq \emptyset$. The inclusion of such a proper
interval $I$ in the unit circle will be denoted as $I\Subset\Seins$. 

The vacuum representation of $\lok{B}$ is given by a map from the set
of proper intervals to \Name{v.Neumann} algebras of bounded operators
on a separable 
\Name{Hilbert} space $\Hilb{H}$ satisfying {\em isotony}, ie for
$I_1\subset I_2\Subset \Seins$ we have
$\lok{B}(I_1)\subset\lok{B}(I_2)$, and {\em locality}, that is: If
$I_1\subset I_2'$, then $\lok{B}(I_1)$ is contained in
$\lok{B}(I_2)'$, the commutant
of  $\lok{B}(I_2)$. It is required as well that there is a
unitary, strongly continuous representation $U$ of the group of global,
chiral, conformal transformations, $\PSL(2,\dopp{R})$, which satisfies
the following: the generator of translations has positive
spectrum (positivity of energy), $U$ implements
the corresponding symmetry 
of $\lok{B}$, ie for $g\in \PSL(2,\dopp{R})$ the adjoint action of
$U(g)$ on local algebras of $\lok{B}$ defines an isomorphism
$\alpha_g$ from any $\lok{B}(I)$ onto the corresponding
$\lok{B}(gI)$, and, finally, $U$ has to contain the trivial representation
exactly once. We choose a vector $\Omega$, the vacuum, of length
$1$ in the corresponding representation space. $\Omega$ has to be
cyclic for $\lok{B}$, which, by the \Name{Reeh-Schlieder} theorem,
amounts to demanding $\lok{B}(I)\Omega$ to be dense in $\Hilb{H}$
for all $I\Subset\Seins$. 

A chiral conformal subtheory $\lok{A}$ embedded in $\lok{B}$, written
as $\lok{A}\subset \lok{B}$, is given by a map from the set of proper
intervals to local \Name{v.Neumann} algebras, $\Seins\Supset I \mapsto
\lok{A}(I)$, with the following properties:
\begin{itemize}
\item {\em Inclusion:} $\lok{A}(I)\subset \lok{B}(I)$ for $I\Subset \Seins$.
\item {\em Isotony:} If $I_1\subset I_2$, then
  $\lok{A}(I_1)\subset\lok{A}(I_2)$.
\item {\em Covariance:} For all $g\in \PSL(2,\dopp{R})$ and
  $I\Subset\Seins$ we have:  $\lok{A}(gI) = \alpha_g
  (\lok{A}(I))$. 
\end{itemize}

These assumptions have a lot of interesting consequences of which we
only name a few directly involved in this work. For instance, $\lok{B}$ has
the {\em {Bisognano-Wichmann} property}, ie the modular data of
the local algebra assigned to the upper half circle have a direct
geometrical interpretation. If the action of scale transformations on
the (chiral) light-ray, which we identify with $\dopp{R}$, reads 
$D(t): x\mapsto e^t x$, $x\in\dopp{R}$, and the modular group of
$\lok{B}(\Seins_+)$ is given by $\Delta^{it}$, then we
have $U(D(-2\pi t))= \Delta^{it}$. Furthermore, the modular
conjugation of $\lok{B}(\Seins_+)$, denoted by $J$, implements the
reflection $x\mapsto -x$. By covariance, this means in particular: the vacuum
representation of $\lok{B}$ satisfies {\em {Haag} duality (on the
circle)}, namely we have  $\lok{B}(I)'=\lok{B}(I')$, $I\Subset\Seins$. 

The local algebras $\lok{B}(I)$, $I\Subset\Seins$, are {\em continuous from
the inside} as well as {\em from the outside}, that is: $\lok{B}(I)$
coincides with the intersection of all local algebras assigned to
proper intervals $J$ containing $\bar{I}$ and is generated by all
its local 
subalgebras (assigned to proper intervals $J$ with
$\bar{J}\subset I$),
respectively. Continuity from the inside
implies {\em weak
additivity}, ie  $\lok{B}(I)$ is generated by the subalgebras
$\lok{B}(J_i)$ for each covering $\bigcup_i J_i = I$ \cite{FJ96}.

The vacuum representation of $\lok{A}$ is contained in the
representation induced by the embedding $\lok{A}\subset\lok{B}$. In
fact, the local inclusions $\lok{A}(I)\subset\lok{B}(I)$, $I\Subset
\Seins$, define a {\em (quantum field theoretical)  net of subfactors} in
the sense of  \Name{Longo and Rehren} \cite{LR95}. By the
\Name{Reeh-Schlieder} theorem, the projection $e_\lok{A}$ onto the
\Name{Hilbert} space resulting 
from the  closure of $\lok{A}(I)\Omega$, $I\Subset\Seins$, does not
depend on $I$ and, because the local subalgebras
$\lok{A}(I)\subset\lok{B}(I)$ are {\em modular covariant} by the
\Name{Bisognano-Wichmann} property of $\lok{B}$ and conformal
covariance of $\lok{A}\subset\lok{B}$, it follows \cite{mT72,vJ83} that for
every $I\Subset \Seins$ we have : $\lok{A}(I)=
\{e_\lok{A}\}'\cap\lok{B}(I)$. For a general summary on modular
covariant subalgebras see eg \cite{hjB97}. 

We denote the \Name{v.Neumann} algebra which is generated by all local
algebras $\lok{A}(I)$, $I\Subset \Seins$, by $\lok{A}$ as well (with a
slight abuse of notation); this algebra contains all local observables
of the theory $\lok{A}$ and all
{\em global} observables associated with the subtheory
$\lok{A}\subset\lok{B}$, that is all bounded operators which are weak
limits of local observables of the subtheory but which are not local
themselves. By the \Name{Borchers-Sugawara} construction \cite{sK02} there is a
unique representation $U^\lok{A}$ of $\PSL(2,\dopp{R})^\sim$, the
universal covering group of $\PSL(2,\dopp{R})$, which consists of
global observables of $\lok{A}$ and implements conformal covariance on
$\lok{A}$ by its adjoint action. Namely, the
representation $U$ factorises as $U\circ\symb{p}(g) =
U^\lok{A}(g)U^{\lok{A}'}(g)$, where $\symb{p}$ denotes the covering
projection form $\PSL(2,\dopp{R})^\sim$ onto $\PSL(2,\dopp{R})$ and
$U^{\lok{A}'}$ is another representation of $\PSL(2,\dopp{R})^\sim$ by
unitaries in $\lok{A}'$. 

In the following there will appear frequently the subgroups of
translations, $T(a)x = x+a$, $x,a\in \dopp{R}$, and of
special conformal transformations, $S(n)x= x/(1+nx)$, $x,n\in
\dopp{R}$. Both groups are inverse-conjugate in $\PSL(2,\dopp{R})$, ie
$T(a)$ is conjugate to $S(-a)$, and the same 
holds true for their images in $\PSL(2,\dopp{R})^\sim$, which we will
denote by $\tilde{T}$ and $\tilde{S}$, respectively.  Furthermore,
we have the groups of rigid conformal rotations, denoted by $R$ and
$\tilde{R}$, respectively, and of dilatations (scale transformations),
$D$, $\tilde{D}$. We adopt the physicists' convention on the
\Name{Lie} algebra and use the same symbols for elements of the
\Name{Lie} algebra and their representatives as elements of an
infinitesimal representation by (essentially) self-adjoint
operators. We use parameters on the three subgroups mentioned so far which
make the subgroup $R$  of rotations naturally isomorphic to
$\dopp{R}/2\pi\dopp{Z}$ and yield the following relation between the
generator of translations, $P$, the generator of special conformal
transformations, $K$, and the generator of rigid conformal rotations,
the conformal \Name{Hamilton}ian  $L_0$: $2L_0=P-K$.  $P$ is a
positive operator iff $L_0$ is positive or, equivalently, iff $-K$ is
positive (eg \cite{sK02}, prop.1). $U^\lok{A}$ and $U^{\lok{A}'}$ both
are of positive energy since $U$ is. Furthermore, both representations
leave the vacuum invariant \cite{sK02} (corollary. 6 and 7).

In the following we deduce, step by step, the sub-geometric character
of the adjoint action of $U^\lok{A}$ (and of $U^{\lok{A}'}$) on
$\lok{B}$. The analysis relies 
 on a single property of the dilatations in $U^\lok{A}$. The notion of
 {\em net-endomorphisms} arises naturally in the course of the argument
 and will be discussed at the end of this section. We, therefore,
 define:

\begin{defi}\label{def:netend}
  $U^\lok{A}$ is said to have the {\bf net-endomorphism property}, if
  the adjoint action of ${U^\lok{A}(\tilde{D}(t))}$, $t\in\dopp{R}$,
  defines a group of automorphisms of  $\lok{B}(\Seins_+)$. 
\end{defi}

This property holds making the following
\begin{ass}\label{ass}
  There is a unitary, strongly continuous, projective representation
  $\Upsilon^\lok{A}$ of the universal covering group of orientation preserving
  diffeomorphisms of the circle, $\Diff_+(\Seins)^\sim$, on $\Hilb{H}$ such
  that:
  \begin{itemize}
  \item If a diffeomorphism $\varphi\in\Diff_+(\Seins)$ is localised in
    $I\Subset\Seins$, ie
    $\varphi\restriction{}{I'}=id\restriction{}{I'}$, it is
    represented 
    by a local observable of $\lok{A}$, namely:
    $\Upsilon^\lok{A}(\symb{p}^{-1}(\varphi))\in\lok{A}(I)$.
  \item
    $\Upsilon^\lok{A}(\tilde{D}(t))U^\lok{A}(\tilde{D}(t))^*
    \in\dopp{C}\Einsop$  
    for all $t\in\dopp{R}$. 
  \end{itemize}
\end{ass}
Here, the covering projection from
$\Diff_+(\Seins)^\sim$ onto $\Diff_+(\Seins)$ is denoted by
$\symb{p}$.  Localised diffeomorphisms $\varphi$ are identified
with their preimage $\symb{p}^{-1}(\varphi)$ in the first
sheet of the covering.

The \Name{Additional Assumption} only
enters through the lemmas \ref{lem:autUA} and \ref{lem:isoUA}, which
we believe to hold true in a lot more general circumstances. It can be
verified in presence of an integrable stress-energy tensor for
$\lok{A}$ (see discussion in appendix). In this case the
representations $\Upsilon^\lok{A}\restriction {\PSL(2,\dopp{R})}^\sim$ and
$U^\lok{A}$ coincide, whereas we have only assumed
that the respective generators agree up to a multiple of $\Einsop$. At
this point we want to stress: We do not assume $\lok{A}$ to be
diffeomorphism covariant, ie the adjoint action of $\Upsilon^\lok{A}$
on $\lok{A}$ to implement a geometric, automorphic action of
$\Diff_+(\Seins)$ on $\lok{A}$.

\begin{lemma}\label{lem:autUA}
  $U^\lok{A}$ has the
  net-\-endomorphism property, if the  \Name{Additional Assumption} holds.
\end{lemma}

\begin{pf}
  By lemma \ref{lem:diffD} there exist, for small $t\in\dopp{R}$,
  diffeomorphisms 
  $g_{\varepsilon}$, $g_{\delta}$ localised in arbitrarily small
  neighbourhoods of $-1$ and  $1$, respectively, and diffeomorphisms
  $g_+$, $g_-$ localised in $\Seins_+$ and $\Seins_-$, respectively,
  such that we have: $D(t)= g_+g_-g_\delta 
  g_\varepsilon$. If the closure of a proper interval $I$ is
  contained in $\Seins_+$,
  we have with an appropriate 
  choice of $g_\delta$, $g_\varepsilon$ by the {\sc Additional Assumption} :
  \begin{equation}\label{eq:autUA}
    U^\lok{A}(\tilde{D}(t)) \lok{B}(I) U^\lok{A}(\tilde{D}(t))^* =
    \Upsilon^\lok{A}(\symb{p}^{-1}(g_+)) \lok{B}(I)
    \Upsilon^\lok{A}(\symb{p}^{-1}(g_+))^*\subset 
    \lok{B}(\Seins_+) \,\, .
  \end{equation}
Because $\lok{B}(\Seins_+)$ is continuous from the inside, we see that
$Ad_{U^\lok{A}(\tilde{D}(t))}$ induces an endomorphism of
$\lok{B}(\Seins_+)$. The same holds true for
$U^\lok{A}(\tilde{D}(-t))$ and, therefore, these endomorphisms are
automorphisms.
\end{pf}

The next step is to give a natural characterisation of one-parameter groups of
unitary operators
which define, by their adjoint action, endomorphism semigroups of a
standard \Name{v.Neumann} algebra.  The following theorem is a
mainly a new formulation of results by \Name{Borchers} and
\Name{Wiesbrock}. Its 
present form is new and appears to be a natural {\em converse of
\Name{Borchers}' theorem on half-sided translations}. The
methods of proof are completely standard, but the result ought to be
made available\footnote{Compare \cite{dD96} for another
characterisation of endomorphism semigroups related to
\Name{Borchers}' theorem.}. 

\begin{satz}\label{bowiesatz}
  Assume $\lok{M}\subset\lok{B}(\Hilb{H})$ to be a
  \Name{v.Neumann} algebra having a cyclic and separating vector
  $\Omega$ in the separable \Name{Hilbert} space $\Hilb{H}$.
  $J,\Delta$ shall stand for the modular data of this pair. Let
  $V(t)$, $t\in\dopp{R}$, be a strongly continuous one-parameter
  group. Then any two from $\{\ref{BWspec},\ref{BWvac},\ref{BWscal}\}$
  imply the remaining two in the list below; $\ref{BWcone}$ yields
  $\ref{BWspec}$, $\ref{BWvac}$, $\ref{BWscal}$.
  \begin{enumerate}
  \item \label{BWspec}
    \begin{enumerate}
    \item \label{BWspecspec}$V(s)=e^{iHs}$, $H\geqslant 0$,
    \item $V(s)\lok{M}V(s)^*\subset\lok{M}$, $s\geqslant 0$.
    \end{enumerate}
  \item \label{BWvac}
    \begin{enumerate}
    \item $V(s)\Omega=\Omega$, $ s\in\dopp{R}$,
    \item $V(s)\lok{M}V(s)^*\subset\lok{M}$, $s\geqslant 0$.
    \end{enumerate}
   \item \label{BWscal}
    \begin{enumerate}
    \item \label{BWscalscal}$\Delta^{it}V(s)\Delta^{-it} = V(e^{-2\pi
      t}s)$, $JV(s)J=V(-s)$, 
      $ t,s\in\dopp{R}$,
    \item $V(s)\lok{M}V(s)^*\subset\lok{M}$, $s\geqslant 0$.
    \end{enumerate}
   \item \label{BWcone}
    \begin{enumerate}
    \item \label{BWcsp}$V(s)=e^{iHs}$, $H\geqslant 0$,
    \item \label{BWcs}$\Delta^{it}V(s)\Delta^{-it} = V(e^{-2\pi t}s)$,
      $t,s\in\dopp{R}$,
    \item \label{BWcc}$\skalar{m_+'\Omega}{V(s)m_+\Omega}\geqslant 0$,
      $s\geqslant 0$, 
      $m_+\in\lok{M}_+$, $m_+'\in\lok{M}'{}_+$.
    \end{enumerate}
   \end{enumerate}
\end{satz}
$\lok{M}_+$ denotes the cone of positive elements in $\lok{M}$,
$\lok{M}_+'$ the cone of positive elements in its commutant $\lok{M}'$.

\begin{pf}
Most of the implications were proved by \Name{Borchers}
and  \Name{Wiesbrock}, respectively:
$\ref{BWspec}\wedge \ref{BWvac} \Rightarrow \ref{BWscal}$:
\cite{hjB92} (cf \cite{mF98}); 
$\ref{BWvac}\wedge \ref{BWscal} \Rightarrow \ref{BWspec}$:
\cite{hwW92}; 
$\ref{BWspec}\wedge \ref{BWscal} \Rightarrow \ref{BWvac}$:
\cite{hjB98a}; 
$\ref{BWspec}\wedge \ref{BWvac} \wedge  \ref{BWscal}\Rightarrow
\ref{BWcone}$: \cite[proposition 2.5.27]{BR87}.  

We prove the remaining statement, namely $\ref{BWcone} \Rightarrow
\ref{BWspec}\wedge \ref{BWvac} \wedge  \ref{BWscal}$, by reduction to 
\cite[theorem 1.1]{hjB97}\footnote{Alternatively, one may use the same
  statement in \cite[theorem 2.5]{hjB97a}.}.  As a first step we look
at the domain of entire analytic vectors with respect to
$\Delta^{iz}$, which we denote by $D_\Delta$, and derive an analytic
continuation of relation $\ref{BWcs}$ as a quadratic form on $D_\Delta$. 
We define: 
\begin{displaymath}
 F(z,w) := 
  \langle\Delta^{i\overline{z}}\psi, e^{ie^{2\pi
        w}H}\Delta^{iz}\phi\rangle \,\, .
\end{displaymath}
According to the spectrum condition on $H$,  $F$ is analytic in $w$ for
$0< Im(w) <\frac{1}{2}$, and this function is bounded and continuous
for the closure of this region; the region itself shall be denoted by
$\dopp{S}$. In fact, by \Name{Hartog}'s theorem,  $F$ is analytic on
$\dopp{C}\times \dopp{S}$ as a function in two complex variables. We
make full  
use of relation $\ref{BWcs}$ by looking at another function $G$, which
agrees with $F$ for $0<
Im(w) + Im(z)< \frac{1}{2}$:  
\begin{displaymath}
  G(z,w) := \langle\psi, e^{ie^{2\pi (w+z)}H}\phi\rangle \,\, .
\end{displaymath}
Evaluating at $w\in\dopp{R}$ and $z=\frac{i}{4}$ we get:
\begin{equation}\label{eq:analcontscaltrans}
   \langle\Delta^{\frac{1}{4}}\psi, e^{ie^{2\pi
        w}H}\Delta^{-\frac{1}{4}}\phi\rangle = \langle\psi,
    e^{-e^{2\pi w}H}\phi\rangle \,\, .
\end{equation}

Both $\psi, \phi$ are of the form
$\psi=\Delta^{-\frac{1}{4}}\psi'$, 
$\phi=\Delta^{\frac{1}{4}}\phi'$, $\psi', \phi'\in D_\Delta$. Since
the set of such $\psi', \phi'$ is dense in $\Hilb{H}$, the equation
above becomes an equation for bounded operators, which yields:
\begin{equation}\label{eq:boskal}
  e^{isH} = \Delta^{-\frac{1}{4}}e^{-sH}\Delta^{\frac{1}{4}}\, , \quad
  s\geqslant 0 \,\, .
\end{equation}

Next, we show invariance of $\Omega$ following arguments from the
proof of \cite[lemma 2.3.c]{hjB98a}: let $E$ be the projection onto
the eigenvectors of
$\Delta$ having eigenvalue $1$. Multiplying the identity
(\ref{eq:boskal}) from both
sides by $E$ leads to:
\begin{displaymath}
  Ee^{isH}E = Ee^{-sH}E\, , \quad  s\geqslant 0\,.
\end{displaymath}
Here, the right hand side is a positive operator and thus we have as well:
\begin{displaymath}
  \klammer{Ee^{isH}E}^* = Ee^{-isH}E = Ee^{-sH}E = Ee^{isH}E\, , \quad
  s\geqslant 0 \,.
\end{displaymath}
According to a standard argument\footnote{Such an argument is given,
  for example, in \cite{sK02} (proof of corollary 7) and uses the
  spectrum condition, the \Name{Phragmen-Lindel\"of} theorem,
  \Name{Schwarz}' reflection principle and \Name{Liouville}'s 
  theorem.}, this invariance with respect to conjugation yields:
 $ Ee^{isH}E = Ee^{i0H}E = E$. 
Therefore, all vectors $\xi$ satisfying $\xi= E\xi$ are 
invariant under the action of $V$ and this means in particular:
$V(s)\Omega = \Omega$, $\forall s\in\dopp{R}$. 

It now follows from $\ref{BWcc}$ and \cite[proposition 2.5.28]{BR87} that 
$e^{-sH}$, $s\geq 0$, leaves the natural cone of
$\klammer{\lok{M},\Omega}$ globally fixed. The other assumptions of
\cite[theorem 1.1]{hjB97} are the identities:
\begin{eqnarray*}
  \Delta^{it}e^{-Hs}\Delta^{-it}&=& e^{-se^{-2\pi t} H}\, , \quad
  s\geqslant 0\,\, ,\\
e^{-Hs} \Omega &=& \Omega\, , \quad s\geqslant 0\,\, .
\end{eqnarray*}
These relations are obvious by analytic continuation of results
derived above. By \cite[theorem 1.1]{hjB97} the adjoint action of
$V(s)$, $s\geq 0$, does indeed induce 
endomorphisms of $\lok{M}$ and we have completed the proof. 
\end{pf}


The arguments in the proof of theorem \ref{bowiesatz} apply, with
minor alterations, to translation groups with negative generator, as eg the
special conformal transformations $U(S(.))$. While $J$ has
the same action, $JU(S(n))J = U(S(-n))$, the scaling behaviour is
opposite:
\begin{equation}
  \label{eq:specconfscal}
  \Delta^{it} U(S(n)) \Delta^{-it} = U(S(e^{2\pi t}n)) \,\, .
\end{equation}
The negative spectrum together with the opposite scaling law
(\ref{eq:specconfscal}) shows that the condition characterising
endomorphism semi-groups is just the same as in condition
$\ref{BWcc}$. Since the arguments are completely analogous as for the case of
positive spectrum and scaling law $\ref{BWscalscal}$, $\ref{BWcs}$ we
state the following corollary without proof:
\begin{folge}\label{cor:bowie}
  The statements in theorem \ref{bowiesatz} still hold, if one
  replaces $\ref{BWspecspec}$, $\ref{BWcsp}$ by $V(s)= e^{iKs}$, $K\leq 0$, and uses
  $\Delta^{it} V(s) \Delta^{-it} = V(e^{2\pi t}s)$, $s,t\in\dopp{R}$,
  instead of 
  $\ref{BWscalscal}$, $\ref{BWcs}$. 
\end{folge}


At this stage our intuition about the geometric impact of $U^\lok{A}$
on $\lok{B}$ can be verified. We will discuss the general situation
right after the following corollary:

\begin{folge}\label{cor:netend}
  Assume ${U^\lok{A}}$ to have the net-endomorphism property. Then 
  the adjoint action of ${U^{\lok{A}'}(\tilde{D}(.))}$ on
  $\lok{B}(\Seins_+)$ defines a group of automorphisms. 

For $s\geq 0$
  the adjoint action of $U^\lok{A}(\tilde{T}(s))$ 
  induces  endomorphisms of $\lok{B}(\Seins_+)$ and the adjoint action
  of $U^\lok{A}(\tilde{T}(-s))$ maps $\lok{B}(\Seins_+)$ into
  $\lok{B}(T(-s)\Seins_+)$. The corresponding statements hold true, if
  one replaces $\lok{A}$ by $\lok{A}'$ or $\tilde{T}(.)$
  by $\tilde{S}(.)$.
\end{folge}

\begin{pf}
The statement on $Ad_{U^{\lok{A}'}(\tilde{D}(.))}$  follows from
$U^{\lok{A}'}= U\!\circ\!\symb{p} \,U^\lok{A}{}^*$ and covariance of
$\lok{B}$. Using the factorisation of $U(T(s))= U^\lok{A}(\tilde{T}(s))U^{\lok{A}'}(\tilde{T}(s))$, covariance and 
isotony of $\lok{B}$,  the
statement on $Ad_{U^{\lok{A}'}(\tilde{D}(.))}$ and invariance of
$\Omega$ with respect to $U^{\lok{A}'}$, we have the following
inequality for all $t\in\dopp{R}$, $s\geq 0$,
$B_+\in\lok{B}(\Seins_+)_+$, $B'_+\in\lok{B}(\Seins_-)_+$:
\begin{eqnarray*}
0 &\leqslant&
\langle
U^{\lok{A}'}(\tilde{D}(t))^*B_+'U^{\lok{A}'}(\tilde{D}(t))\Omega,\, 
  U(T(s))
  U^{\lok{A}'}(\tilde{D}(t))^*B_+U^{\lok{A}'}(\tilde{D}(t))\Omega\rangle\\ 
&=& \langle B_+'\Omega,\,
  U^\lok{A}(\tilde{T}(s))U^{\lok{A}'}(\tilde{T}(e^{t}s))
 B_+\Omega\rangle \,\, .
\end{eqnarray*}
In the limit $t\rightarrow -\infty$ strong continuity of
$U^{\lok{A}'}$ implies $\langle B_+'\Omega, 
  U^\lok{A}(\tilde{T}(s)) B_+\Omega\rangle \geq 0$, which in turn yields the
statement on $U^\lok{A}(\tilde{T}(s))$, $s\geq 0$, by theorem
\ref{bowiesatz}. Following
the same argument 
with $\lok{A}$ instead of $\lok{A}'$ and vice versa leads to the
corresponding statement on  $U^{\lok{A}'}(\tilde{T}(s))$, $s\geq
0$. If one replaces in both statements $\tilde{T}(s)$ by
$\tilde{S}(s)$, one may apply the argument as well, but using the
limit $t\rightarrow \infty$ and corollary \ref{cor:bowie}.

The remainder follows immediately from the following argument, which
we indicate for the translations represented through $U^\lok{A}$:
\begin{displaymath}
  Ad_{U^\lok{A}(\tilde{T}(-s))} \lok{B}(\Seins_+) = Ad_{U(T(-s))}
Ad_{U^{\lok{A}'}(\tilde{T}(s))}\lok{B}(\Seins_+)
\subset \lok{B}(T(-s)\Seins_+) \,\,. 
\end{displaymath}
\end{pf}


The geometric impact of a general $U^\lok{A}(\tilde{g})$,
$\tilde{g}\in \PSL(2,\dopp{R})^\sim$, on an arbitrary local algebra
$\lok{B}(I)$ is discussed easily. We may restrict our attention to
group elements $\tilde{g}$ for which there is a single sheet of the
covering projection $\symb{p}$ containing both $\tilde{g}$ and the
identity, as the following discussion indicates. 

Every element $g$ in $\PSL(2,\dopp{R})$ is contained in (at least) one
one-parameter group\footnote{I am indebted to D. Guido for providing
  the reference. In the particular case of $\PSL(2,\dopp{R})$  this
  fact may be checked directly (cf \cite{sK03d}).}
\cite{mM94,mM97}.  We use the local identification of
one-parameter subgroups in $\PSL(2,\dopp{R})$ and in
$\PSL(2,\dopp{R})^\sim$, choose a parametrisation such 
that $\tilde{g}=\tilde{g}(1)$, $id= \tilde{g}(0)$, and we set
$\gamma_{\tilde{g}}(I) := \bigcup_{\tau=0}^1 \symb{p}(\tilde{g}(\tau))
I$. For $\tilde{g}$ further away from the identity we set
$\gamma_{\tilde{g}}(I)=\Seins$ and take $\lok{B}(\Seins)$ to be the
algebra of all bounded operators on $\Hilb{H}$.  Then we have:
  
\begin{prop}\label{prop:netend}
Assume $U^\lok{A}$ to have the net-endomorphism property. Then we have   
for any $\tilde{g}\in \PSL(2,\dopp{R})^\sim$ and any $I\Subset
  \Seins$: $Ad_{U^\lok{A}(\tilde{g})}\lok{B}(I)\subset
  \lok{B}(\gamma_{\tilde{g}}(I))$,  and
  $Ad_{U^{\lok{A}'}(\tilde{g})}\lok{B}(I)\subset 
  \lok{B}(\gamma_{\tilde{g}}(I))$.
\end{prop}

\begin{pf}
Each proper interval $I$ in $\Seins$ may be
identified by the ordered 
pair consisting of its boundary points,  $z_+$
and $z_-$.
We define three one-parameter subgroups in $\PSL(2,\dopp{R})$
referring to each $I\Subset\Seins$ with
respect to a particular choice $h\in\PSL(2,\dopp{R})$ satisfying
$h\Seins_+ =I$: $D_I(.) = h D(.) h^{-1}$, $T_I(.) = h T(.)
h^{-1}$, $S_I(.) = h S(.) h^{-1}$.

Each element $g$ in $\PSL(2,\dopp{R})$ is fixed, up to a dilatation
$D_I(t)$, by its action on $\{z_+,z_-\}$. Under the action of elements
$g(\tau)$, $\tau=0,\ldots, 1$, interpolating in the one-parameter
group associated with $g$ between the identity and $g=g(1)$, the orbits
of $z_\pm$ are given by monotonous functions $z_+(\tau)$,
$z_-(\tau)$.   Demanding $s$, $n$,
$t$ to depend continuously on $\tau$ and to take value $0$ at $\tau=0$,
every $g(\tau)$ may be represented as
 $g(\tau)=
S_{T_I(s(\tau))I}(n(\tau)) T_I(s(\tau)) D_I(t(\tau))$ or as $g(\tau)=
T_{S_I(n(\tau))I}(s(\tau)) S_I(n(\tau)) D_I(t(\tau))$. We choose one
form which works for all interpolating elements. By the requirements we
have made it is ensured that the representation works (after obvious
identifications) in $\PSL(2,\dopp{R})^\sim$ as well.  Corollary
\ref{cor:netend} implies the claim of the proposition now.
\end{pf}

This proves in particular: For every $I\Subset\Seins$ there is a
neighbourhood of the identity in $\PSL(2,\dopp{R})^\sim$ for which
the action of $Ad_{U^\lok{A}(.)}$ on $\lok{B}(I)$   delivers local
observables. 

We have found $Ad_{U^\lok{A}}$ to induce homomorphisms from local
algebras of $\lok{B}$ into algebras associated with an enlarged localisation
region. This
sub-geometrical action respects isotony, ie the net-structure. The
adjoint action of $U$ induces the covariance isomorphisms of local
algebras and one 
usually regards these as {\em automorphisms of the net $\lok{B}$}. We
consider, therefore, the term {\em net-endomorphisms} appropiate. The
automorphic action of $Ad_{U^\lok{A}(\tilde{D}(.))}$ on 
$\lok{B}(\Seins_+)$ which we proved in lemma \ref{lem:autUA} does not,
apparently, follow from the endomorphism property for the translation
subgroups in corollary \ref{cor:netend}. This motivated definition
\ref{def:netend} above.

In the next section we give a {\em holographic} interpretation
of the  net-en\-do\-mor\-phism property. This shows that the results
achieved so far are satisfactory and yield an interesting and useful
new insight into structures associated with chiral conformal subnets
and their \Name{Coset} models. 

 
\section{Chiral holography}
\label{sec:chihol}

The mapping $(\tilde{g},\tilde{h})\mapsto U^\lok{A}(\tilde{g})
U^{\lok{A}'}(\tilde{h})$ defines 
 a representation $U^\lok{A}\times U^{\lok{A}'}$ of
 the group  $\PSL(2,\dopp{R})^\sim\times
 \PSL(2,\dopp{R})^\sim$. This is, in fact, 
a representation 
of the conformal symmetry group of a local conformal quantum theory in
$\opo$ dimensions, which is isomorphic to
$(\PSL(2,\dopp{R})^\sim\times \PSL(2,\dopp{R})^\sim)/\dopp{Z}$. This
factor group arises, if one identifies the simultaneous rigid
conformal rotation by $2\pi$, namely
$(\tilde{R}(2\pi),\tilde{R}(2\pi))$, with the trivial
transformation. The last section taught us a lot about the
sub-geometrical action of $U^\lok{A}$, $U^{\lok{A}'}$ on
the local observables in $\lok{B}$. So, it is natural to look for a
relation between the geometrical character of this action and structures in
$\opo$ dimensions.

This relation turns out to be a complete correspondence: We construct
a $\opo$-dimensional, local, conformal theory from the 
original chiral theory $\lok{B}$ applying the net-endomorphism
property of $U^\lok{A}$. In
order to prove locality in $\opo$ dimensions we are led to a
particular choice of light-cone coordinates, by which  the
original local algebras $\lok{B}(I)$, $I\Subset\Seins$, are included
in the $\opo$-dimensional picture as {\em time zero algebras}. This
choice of coordinates yields an unphysical spectrum condition:
translations in the right {\em spacelike} wedge 
have positive spectrum. Whereas this prohibits an interpretation of
the new theory as a genuinely physical one, where we would have
positivity of the spectrum in future-like directions, the construction
does provide us with a useful geometrical picture for
questions concerned with chiral subnets and their \Name{Coset} models.
For this reason we regard the result of our construction as a
local, conformal {\em quasi-theory} in $\opo$ dimensions.

If, on the opposite, one takes a (physical) conformal quantum theory
in $\opo$ dimensions and defines a chiral conformal net by restriction
to time zero algebras, a similar phenomenon arises (cf
\cite{KLM01,rL01}): the spectrum condition disappears altogether, but
powerful tools of local quantum theory are available still, because
the \Name{Reeh-Schlieder} property survives. In our case there remains a spectrum condition from
which one can still derive the \Name{Reeh-Schlieder} property. In
this sense we find a natural ``converse'' of the restriction process
which  justifies the term {\em chiral holography} for our
construction. 

The main result of this section will be proved by making contact with
the analysis of \Name{Brunetti, Guido and Longo} \cite{BGL93} who
discussed conformal quantum field theories in general spacetime
dimensions 
as local quantum theories on the conformal covering of the respective
\Name{Minkowski} space as extensions of local nets living on
\Name{Minkowski} space itself.

In $\opo$ dimensions, \Name{Minkowski} space $\dopp{M}$ is the
Cartesian product of two chiral light-rays, which we take as
light-cone coordinates of  $\dopp{M}$. One
arrives at the (physical) conformal covering $\widetilde{\dopp{M}}$ of
$\dopp{M}$, if one compactifies both light-rays adding the points at
infinity, takes the infinite, simply connected covering of the compactification
$\Seins\times\Seins$, which yields $\dopp{R}\times\dopp{R}$,
and, finally, one identifies all points which are connected by the
action of simultaneous rigid conformal rotations by $2\pi$. The result
has the shape of a cylinder having infinite timelike
extension: $\widetilde{\dopp{M}}= \Seins\times\dopp{R}$. Without the
final identification we would have 
spacelike separated copies of $\dopp{M}$ in covering space, which we
consider unphysical; conformally covariant quantum fields can be
proven to live on this (physical) conformal covering of
\Name{Minkowski} space, see \cite{LM75}. 

Light-rays in $\widetilde{\dopp{M}}$ are infinitely extended, universal
coverings of the compactified light-rays and serve well as light-cone
coordinates. The localisation regions are
{\em $\opo$-dimensional double cones} given as Cartesian product
of two intervals, $I\times J$, where $I$, $J$ are properly contained
in a single copy of $\Seins$ on the left and right light-rays,
respectively, in $\widetilde{\dopp{M}}$. 

$\PSL(2,\dopp{R})^\sim$  has an action on the
infinite covering $\dopp{R}$ of $\Seins$ which is transitive
for the intervals properly contained in a single copy of
$\Seins$. We exclude the point of
infinity from $\Seins$ and choose a fixed interval $I$ which is
properly contained in the remainder. This interval is identified with
its first (pre-)image in covering space and we choose for any pair of proper
intervals $J_{L,R}$ group elements 
$\tilde{g}_{L,R}\in \PSL(2,\dopp{R})^\sim$ satisfying
$J_{L}=\tilde{g}_{L}I$, 
$J_R:=\tilde{g}_RI$. Making use of this choice we  define a set of
(local) algebras indexed by $\opo$-dimensional double cones:
\begin{equation}
  \label{eq:oodef}
  \lok{B}^{1+1}(J_L\times J_R) := U^\lok{A}(\tilde{g}_{L})
  U^{\lok{A}'}(\tilde{g}_{R}) \lok{B}(I) U^\lok{A}(\tilde{g}_{L})^*
  U^{\lok{A}'}(\tilde{g}_{R})^* \,\,. 
\end{equation}
By covariance of $\lok{B}$, the resulting algebra $\lok{B}^{1+1}(J_L\times
J_R)$ is uniquely determined by $J_L\times J_R$.   

Furthermore, we define a covering projection $\symb{p}$ from
$\dopp{R}$ onto $\Seins$ referring to the covering projection
$\symb{p}: \PSL(2,\dopp{R})^\sim\rightarrow \PSL(2,\dopp{R})$ such that
we have: $\symb{p} J_{L,R} := \symb{p}(\tilde{g}_{L,R}) I$.  This
definition enables us to state two identities for the 
algebras defined in equation (\ref{eq:oodef}):
\begin{eqnarray*}
  \lok{B}^{1+1}(J_L\times J_R) &=& U^\lok{A}(\tilde{g}_{L}\tilde{g}_{R}{}^{-1})
  \lok{B}(\symb{p} J_{R})
  U^\lok{A}(\tilde{g}_{L}\tilde{g}_{R}{}^{-1})^*\\
&=&  U^{\lok{A}'}(\tilde{g}_{R}\tilde{g}_{L}{}^{-1}) \lok{B}(\symb{p} J_{L}) 
  U^{\lok{A}'}(\tilde{g}_{R}\tilde{g}_{L}{}^{-1})^* \,\,. 
\end{eqnarray*}

Double cones $J\times J$, which are centered at the time zero axis, are
called {\em time zero double cones} and we get for the corresponding
{\em time zero algebras}: 
  $
\lok{B}^{1+1}(J\times J) = \lok{B}(\symb{p} J)
$.
Thus, the local algebras of the original chiral conformal theory
$\lok{B}$ are included into the new quasi-theory $\lok{B}^{1+1}$ as time zero
algebras. Now we are prepared to state the main result of
this section:
\begin{satz}\label{th:chihol}
  If $\lok{A} \subset\lok{B}$ is an inclusion of chiral conformal
  theories and if the unique inner-implementing representation
  $U^\lok{A}$ associated with this inclusion has the net-endomorphism
  property, then equation (\ref{eq:oodef}) defines a set
  $\lok{B}^{1+1}$ of local 
  algebras assigned to double cones in $\opo$-dimensional conformal
  space time, $\widetilde{\dopp{M}}$, having all but one of the usual
  properties of a local, conformal, weakly additive quantum theory
  in $\opo$ dimensions (see \cite{BGL93}): the spectrum condition holds
  for translations in the right spacelike wedge.
\end{satz}

\begin{pf}
Obviously, the set $\lok{B}^{1+1}$ of local algebras is covariant with
respect to the representation $U^\lok{A}\times U^{\lok{A}'}$. Because
of the identity
$U^\lok{A}(\tilde{R}(2\pi))U^{\lok{A}'}(\tilde{R}(2\pi))=\Einsop$
the set $\lok{B}^{1+1}$ is in fact labelled by the double cones in
$\widetilde{\dopp{M}}$ and $U^\lok{A}\times U^{\lok{A}'}$ is a
representation of the conformal group in $\opo$ dimensions, namely the
group $(\PSL(2,\dopp{R})^\sim\times
\PSL(2,\dopp{R})^\sim)/\dopp{Z}$. The spectrum condition for
$U^\lok{A}\times U^{\lok{A}'}$ was proved in \cite{sK02}(corollary 6).

The vacuum vector is invariant with respect to $U^\lok{A}\times
U^{\lok{A}'}$ \cite{sK02}(corollary 7) and it is a basis for the space
of vectors with this property, because the space of $U$-invariant
vectors is one dimensional. $\Omega$ is cyclic for all local algebras
in $\lok{B}^{1+1}$ because of the \Name{Reeh-Schlieder} property of
$\lok{B}$. 

Isotony follows directly from the net-endomorphism property. An
inclusion of $\opo$-dimensional double cones $\tilde{g}_LI\times
\tilde{g}_RI\subset \tilde{h}_LI\times
\tilde{h}_RI$ contained in \Name{Minkowski} space $\dopp{M}$, yields
the relations: $\tilde{h}_{L,R}{}^{-1}\tilde{g}_{L,R}I\subset
I$. Applying proposition \ref{prop:netend} we get:
$Ad_{U^{\lok{A}'}(\tilde{h}_{R}{}^{-1}\tilde{g}_{R})U^\lok{A}(\tilde{h}_{L}{}^{-1}\tilde{g}_{L})}\lok{B(I)}\subset
\lok{B}(I)$. This is equivalent to $\lok{B}^{1+1}(\tilde{g}_LI\times
\tilde{g}_RI)\subset \lok{B}^{1+1}(\tilde{h}_LI\times
\tilde{h}_RI)$. 

Locality for double cones in $\dopp{M}$ is shown easily as well. We
can reduce the  discussion to  the situation where there is a double
cone $J_1\times J_2$ spacelike to our basic time zero double cone $I\times I$
simply by applying an appropriate transformation. There is a time zero
double cone $J\times J$ which contains $J_1\times J_2$ and is spacelike to
$I\times I$. Since we have shown isotony for $\lok{B}^{1+1}$, locality for
this set follows from locality of $\lok{B}$.    

Weak additivity may be proved as in the chiral case. By scale
covariance the local algebras of $\lok{B}^{1+1}$ are continuous from
the inside as well as from the outside \cite{LRT78}. Because we can
restrict the discussion to time zero algebras and the argument of
\Name{J\"{o}r\ss} \cite{mJ96} for the corresponding chiral situation
may be extended directly, we have weak additivity for
$\lok{B}^{1+1}$. 

The proof is complete, if one recognises that the proof of
\cite{BGL93} (proposition 1.9, on the unique extendibility of
$\lok{B}^{1+1}$ from $\dopp{M}$ to all of $\widetilde{\dopp{M}}$) only
requires the prerequisites 
established so far. In particular, not the spectrum condition itself is
needed, but only its consequence, the \Name{Reeh-Schlieder} property. 
\end{pf}

In light of this theorem we obtain a straightforward interpretation
of the sub-geometrical action of $U^\lok{A}$ on $\lok{B}$. If we apply
a chiral coordinate transformation $\tilde{g}_R$ to a time zero double 
cone $J\times J$ and if we test the localisation of the correspondingly
transformed local algebra of $\lok{B}^{1+1}$ only by looking at time
zero algebras, then we find that the result commutes just with
 time zero algebras $\lok{B}(K)$ assigned to
proper intervals $K$ contained in the causal complement of
$\gamma_{\tilde{g}_R}J$. The statement of proposition
\ref{prop:netend} follows from \Name{Haag} duality of $\lok{B}$. 

The theorem has some direct applications to chiral subtheories and
their \Name{Coset} models: We have found that the maximal \Name{Coset}
model $\lok{C}_{max}$ associated with a subtheory
$\lok{A}\subset\lok{B}$ may be regarded as the chiral conformal theory
of all right chiral observables in  $\lok{B}^{1+1}$ in the sense of
\Name{Rehren} \cite{khR00}, ie the local observables of
$\lok{B}^{1+1}$ which are invariant under the action of
transformations on the left light-cone coordinate only.

The observables of $\lok{A}$ may be viewed as left chiral
observables and the chiral conformal subnet
$\lok{A}_{max}\subset\lok{B}$ consisting of local observables
invariant with respect to the action of $U^{\lok{A}'}$ (and
hence covariant with respect to the action of $U^\lok{A}$) is to
be identified with the chiral theory of all left chiral observables in
$\lok{B}^{1+1}$. 

Thus, we have identified $\lok{A}_{max}$ and $\lok{C}_{max}$ as
fixed-points of a space-time symmetry acting on a
suitably extended theory, namely $\lok{B}^{1+1}$. In presence of the
net-endomorphism property it is not necessary to extend the
``classical''  symmetry concept (see eg \cite{hA92}), if one wants to
interpret the chiral subtheories $\lok{A}_{max}$ and $\lok{C}_{max}$
as fixed-points of a symmetry; all one has to do is to extend the
theory $\lok{B}$ to its holographic image. Generalisations of the
symmetry concept are necessary for a large class of chiral conformal
subtheories \cite{LR95, khR94a}. 

Further remarks\footnote{For further details see
  \cite{sK03d} and the appendix.}: Another interesting, direct
consequence of theorem 
\ref{th:chihol} is the following: The cyclic subspaces of
$\lok{C}_{max}$ and $\lok{A}_{max}$, namely
$\overline{\lok{C}_{max}(I)\Omega}$ and
$\overline{\lok{A}_{max}(I)\Omega}$, coincide
with the spaces of $U^\lok{A}$- and $U^{\lok{A}'}$-invariant vectors,
respectively\footnote{By \cite{khR00} (lemma 2.3). The proof of
  proposition \ref{prop:lemma23} includes  an
  alternative argument leading to this statement.}. By this, results from
character arguments on inclusions of current algebras have a direct
and rigorous meaning to the analysis of the respective inclusions of
chiral conformal theories and \Name{Coset} models. 

Here one starts with an inclusion of current algebras,
$\lok{A}\subset\lok{B}$, and looks at the decomposition of the vacuum
representation of $\lok{B}$ when restricted to the subtheory
$\lok{A}\vee\lok{C}\subset\lok{B}$, $\lok{C}$ some \Name{Coset} model
associated with $\lok{A}\subset\lok{B}$. \Name{Goddard, Kent and
  Olive} \cite{GKO86} constructed the minimal series of the
\Name{Virasoro} algebra, ie the quantum field theories generated by
the stress-energy tensors having central charge less than $1$, as
\Name{Coset} models associated with the inclusion of current
algebras $SU(2)_{k+1}\subset SU(2)_1\otimes SU(2)_k$, $k=1, 2,
\ldots$. Their decomposition formulae show the \Name{Coset} stress-energy
tensor to generate all of $\lok{C}_{max}$ and
$SU(2)_{k+1}$ to coincide with its maximal covariant
extension\footnote{\Name{Kawahigashi and Longo} gave an
  alternative argument on this point  \cite{KL02} (lemma 3.2,
  corollary 3.3).},
$\lok{A}_{max}$. We call the chiral conformal quantum theories
generated by a stress-energy tensor with central charge $c$ less than
one ``$\lok{V}ir_{c<1}$ models''. 

\Name{Kac and Wakimoto} \cite{KW88} gave a list of such
decompositions for inclusions $\lok{A}\vee\lok{C}\subset\lok{B}$,
where $\lok{A}\subset\lok{B}$ is an inclusion of current algebras and
$\lok{C}$ is $\lok{V}ir_{c<1}$-\Name{Coset} model associated with this
inclusion. Their list
includes some examples in which $\lok{A}_{max}$ and/ or
$\lok{C}_{max}$ are non-trivial local extensions of $\lok{A}$ and $\lok{C}$,
respectively. 

The local extensions of all $\lok{V}ir_{c<1}$ models have been
classified completely by \Name{Kawahigashi and Longo}
\cite{KL02}. Most of the non-trivial ones are given by {\em orbifolds}:
the local extension contains the  $\lok{V}ir_{c<1}$ model as
fix-point subtheory with  respect to a $\dopp{Z}_2$ symmetry. Some of
these are among the examples of \cite{KW88}. Only
four local extensions are of a different type. For two of these
\Name{Kawahigashi and Longo} gave a rigorous interpretation as
\Name{Coset} models following suggestions of \Name{B\"{o}ckenhauer and
  Evans} \cite{BE99a}. 

One of the remaining two is given as a maximal \Name{Coset} model by
chiral holography 
and the results of \cite{KW88}: the vacuum representations of the
maximal \Name{Coset} models  associated with the current algebra
inclusions $SU(9)_2\subset E(8)_2$ and $E(8)_3\subset E(8)_2\otimes
E(8)_1$ both decompose upon restriction to the
$\lok{V}ir_{c=\frac{21}{22}}$ model into the direct sum of the vacuum
representation and the representation with highest weight
$8$. Following \Name{Kawahigashi and Longo} there is
only one local extension with this decomposition, namely the extension
$(A_{10}, E_6)$ according to the classification scheme
\cite{KL02}, which thus is identified as the maximal  \Name{Coset}
model associated with both current algebra inclusions. 

By classification results on inclusions of current algebras
giving rise to $\lok{V}ir_{c<1}$-\Name{Coset} models \cite{BG87},
the forth exceptional local extension, namely $(A_{28},E_8)$ of
$\lok{V}ir_{c=\frac{144}{145}}$, does not seem to be available by a
\Name{Coset} construction using current algebra inclusions. However,
the local extension is known to exist by an abstract construction
relying on the \Name{DHR}-data of  $\lok{V}ir_{c=\frac{144}{145}}$
\cite{KL02} and thus appears to be a genuine achievement of local
quantum physics.

As we have mentioned before, the holographic image $\lok{B}^{1+1}$ may
not be interpreted as a physical model because of its peculiar
spectrum condition. The picture changes in this aspect, if the
net-endomorphism property takes a sharper form, namely if the transformed
algebra, $Ad_{U^\lok{A}(\tilde{g})}\lok{B}(I)$,
commutes with all $\lok{B}(J)$, $J$ a proper interval contained in
$I'\cap(\symb{p}(\tilde{g})I)'$. After transfer into the holographic
picture this property can easily be seen to be equivalent to
{\em timelike commutativity}
 of $\lok{B}^{1+1}$.  In this case we may interchange the role of
 space and time 
 in the holographic picture and get a physically sensible conformal
 quantum theory in $\opo$ dimensions. 

It is not difficult to extend arguments of \Name{Longo} \cite{rL01} on
chiral subtheories to the $\opo$-dimensional inclusions
$\lok{A}_{max}\vee\lok{C}_{max}\subset\lok{B}^{1+1}$: One can show
that the representation of  $\lok{A}_{max}\otimes\lok{C}_{max}$ induced
by the inclusion is unitarily equivalent to a localised representation
$\rho$. In case $\rho$ has finite statistical
dimension and a finite decomposition into tensor products
$\sigma_i\otimes \tau_j$ of irreducible localised representations 
$\sigma_i$ of $\lok{A}_{max}$ and $\tau_j$ of $\lok{C}_{max}$,
respectively, this can be applied to the situation where
$\lok{B}^{1+1}$ fulfills timelike commutativity in order to derive a
necessary criterion for this particular property to hold.

One knows that the statistical
phases of $\sigma_i$ and $\tau_j$ in a tensor product $\sigma_i\otimes
\tau_j$ occurring in $\rho$ have to be conjugate to each other because
of spacelike commutativity \cite{khR01} (corollary 3.2.). The argument
covers the situation of timelike commutativity, where it
forces the same statistical phases to coincide. The
conformal spin-statistics theorem \cite{GL96} tells us then that
$U^\lok{A}(\tilde{R}(2\pi))$ has to have spectrum in $\{\pm 1\}$,
ie the conformal highest weights associated with $\sigma_i$ and
$\tau_j$ have to lie in $\frac{1}{2} \dopp{N}$. 

This necessary condition excludes all inclusions of current algebras
known to the author (except the ones that one can make up
trivially). The result on the conformal highest weights is well known
for (quasi-) primary fields in $\opo$ dimensions commuting with themselves
not only for spacelike, but also for timelike separation. The analyticity
properties of their two-point function force both chiral scaling
dimensions of such fields to be half integers.
  
\section{Isotony problem}
\label{sec:isoprob}

In this section we use the  \Name{Additional Assumption}  
in order  to  solve the isotony  problem for the local relative
commutants $\lok{C}_I$ of an inclusion  of chiral conformal theories,
$\lok{A}\subset\lok{B}$. Once their isotony is proved, they are known
to coincide with the  local
algebras of the maximal \Name{Coset} model $\lok{C}_{max}$ associated
with $\lok{A}\subset \lok{B}$. This way, we
reach the main goal of this paper: the maximal \Name{Coset} model is
found to be of a local nature, ie it is determined completely by local
data. 
  
As we mentioned in the introduction, the isotony problem requires a
discussion using an argument suited for  our specific scenario. We
reduce the task by a purely group theoretical lemma first, which is a
slightly extended version of a result of \Name{Guido and
  Longo} \cite{GL96}. In a side remark we use this lemma to
characterise the subnets $\lok{A}_{max}$ and $\lok{C}_{max}$ and the
vacuum subrepresentation of $\lok{A}_{max}\vee\lok{C}_{max}$. The argument
is continued by a proposition providing necessary and sufficient
conditions for isotony of local relative commutants to hold. Aside of
being an intermediate step of our analysis, it illustrates the
 character of the isotony problem.  The argument is
completed by an application  of the  \Name{Additional Assumption}  and
summarised in the main theorem  of this work. In the remainder   of
this section we make some remarks on  immediate applications of the
theorem and relations to other works.                          

\begin{lemma}\label{lem:invvec}
$\Hilb{H}$ a separable \Name{Hilbert} space, $V$ a unitary, strongly
continuous representation  of $\PSL(2,\dopp{R})^\sim$  on
$\Hilb{H}$. If $H\subset \PSL(2,\dopp{R})^\sim$ is a subgroup having
closed, non-compact image in  $\PSL(2,\dopp{R})$    under the action of the
covering projection $\symb{p}$, then each $V|_{H}$-invariant vector is
in fact $V$-invariant. If $V$ is a representation of positive energy,
then each vector which is invariant with respect to $V(\tilde{R}(.))$
is $V$-invariant as well.  
\end{lemma}
\begin{pf}
The proof of the claim is, up to trivial modifications, identical to
the one indicated by 
\Name{Guido and Longo} for \cite{GL96}, corollary  B.2. For the
reader's convenience we include a sketch of the argument. 

First, one recognises that it is completely sufficient to discuss the
complement of the trivial subrepresentation, $V^\perp$, on
 the \Name{Hilbert} subspace $\Hilb{H}^\perp$, which
contains no vectors invariant with respect to the 
whole of $V$. We  decompose $V^\perp$ into a direct integral of
irreducible representations  $V_x$.

We look at $V_x\otimes\overline{V_x}$, which can easily be seen not to
contain the trivial representation, because $V_x$ is infinite
dimensional (cf eg \cite{dG93}). Moreover, 
$V_x\otimes\overline{V_x}$ is a representation of $\PSL(2,\dopp{R})$. 
Now we are in the position to
 apply  \cite{rZ84} 
(theorem 2.2.20) and thus we have for any
$\xi_x\in\Hilb{H}_x$:          
\begin{displaymath}
  \lim_{\symb{p}(\tilde{g})\rightarrow \infty}
  \betrag{\skalar{V_x(\tilde{g})\xi_x}{\xi_x}}^2 =
  \lim_{\symb{p}(\tilde{g})\rightarrow \infty}
  \skalar{V_x(\tilde{g})\otimes\overline{V_x}(\tilde{g})
  \xi_x\otimes\overline{\xi_x}}{\xi_x\otimes\overline{\xi_x}}  
  = 0 \,\,. 
\end{displaymath}
If we apply this to a $V_x|_H$-invariant vector $\psi_x$, we readily
see: $\psi_x=0$. Integrating over $x$ yields the first statement of
the lemma. 

The result on rigid conformal rotations may be deduced in the same
manner: The irreducible representations $V_x$ are almost all of
positive energy and the only irreducible representation of
$\PSL(2,\dopp{R})^\sim$ having positive energy and containing a
non-trivial $\tilde{R}(.)$-invariant vector is the trivial
representation \cite{dG93}.                 
\end{pf}

The following result is partly known from \cite{khR00}
(lemma 2.3); we give an alternative proof here. Together with the
other parts, this proposition may be viewed as a generalised version
of \cite{fX00} (theorem 2.4), which is formulated for a particular
class of chiral subnets.
\begin{prop}\label{prop:lemma23}
  Assume $U^\lok{A}$ to have the net-endomorphism property and denote
  the projections onto the subspaces of $U^\lok{A}$- and
  $U^{\lok{A}'}$-    
  invariant vectors by $E_\lok{A}$ and $E_{\lok{A}'}$,
  respectively. Then we have for the maximal $U^\lok{A}$-covariant
  extension of $\lok{A}$,  given by
  $\lok{A}_{max}(I):=\{U^{\lok{A}'}\}'\cap\lok{B}(I)$, and  the
  maximal \Name{Coset} model associated with $\lok{A}\subset\lok{B}$,
  given by $\lok{C}_{max}(I):=\{U^{\lok{A}}\}'\cap\lok{B}(I)$, for
  arbitrary $I \Subset\Seins$: 
  \begin{equation}
    \label{eq:lemma23}
  \overline{\lok{A}_{max}(I)\Omega}=  E_{\lok{A}'}\Hilb{H}\, , \quad
  \overline{\lok{C}_{max}(I)\Omega}= 
  E_{\lok{A}}\Hilb{H} \,\, . 
  \end{equation}
For any \Name{Coset} model $\lok{C}$ associated with
  $\lok{A}\subset\lok{B}$ we have a unitary equivalence of chiral
  conformal theories:
  $\lok{A}\vee\lok{C}e_{\lok{A}\vee\lok{C}}  \cong
  \lok{A}e_\lok{A}\otimes\lok{C}e_\lok{C}$. $E_{\lok{A}}\Hilb{H}$ has a direct
  interpretation as multiplicity space of the vacuum subrepresentation
  of $\lok{A}\subset\lok{B}$.  
\end{prop}
\begin{pf}
  Concerning the proof of (\ref{eq:lemma23}) we may restrict to
  $I=\Seins_+$ (because of the \Name{Reeh-Schlieder} theorem). By
  lemma \ref{lem:invvec} the spaces of vectors which 
  are invariant with respect to translations are identical with
  $E_\lok{A}\Hilb{H}$ and $E_{\lok{A}'}\Hilb{H}$, respectively. Taking
  into account corollary \ref{cor:netend} above the statement
  (\ref{eq:lemma23}) was
  proved by \Name{Borchers} \cite{hjB98a} (theorem 2.6.3).

    Straightforward verification shows
  $\lok{A}e_\lok{A}\otimes\lok{C}e_\lok{C}$ to be a chiral conformal
  theory with the obvious definitions: its vacuum is given by
  $\Omega\otimes\Omega$, the representation implementing covariance is
  $Ue_\lok{A}\otimes Ue_\lok{C}(.)$, its representation space is
  $e_\lok{A}\Hilb{H}\otimes e_\lok{C}\Hilb{H}$.  The factoriality of
  the local algebras proves that $\Omega\otimes\Omega$ is (up to
  scalar multiples) unique  \cite{GL96}(proposition 1.2),  \cite{mT79}
  (IV.5., corollary 
  5.11). One can  establish uniqueness 
  of $\Omega\otimes\Omega$ by group theoretic arguments as in
  lemma \ref{lem:invvec} as well.
  
  We now look at the restrictions of
  $\lok{A}\vee\lok{C}e_{\lok{A}\vee\lok{C}}$ and
  $\lok{A}e_\lok{A}\otimes\lok{C}e_\lok{C}$ to the chiral light-ray,
  $\dopp{R}$. $\Omega$ is separating for $\bigcup_{I\Subset\dopp{R}}
  \lok{A}\vee\lok{C}e_{\lok{A}\vee\lok{C}}(I)$, the union of all
  local algebras assigned to compact intervals in $\dopp{R}$. Thus, we
  are allowed to define a linear operator $W$ densely by: 
  \begin{equation}
    \label{eq:defcospauni}
    W A C \Omega := A\Omega\otimes C\Omega\, , \,\, A\in \lok{A}(I),
    C\in \lok{C}(I), I\Subset\dopp{R} \,\, .
  \end{equation}

  The vacuum is a product state for 
  $\bigcup_{I\Subset\dopp{R}}
  \lok{A}\vee\lok{C}e_{\lok{A}\vee\lok{C}}(I)$ (a corollary to
  \Name{Takesaki}'s  theorem on modular covariant subalgebras
  \cite{mT72}). Hence,  $W$ is bounded and extends by continuity to an
  isometry, as one may  readily verify. Moreover, it is elementary to
  check that $WW^*$ and $W^*W$ commute with the respective  restricted nets
  on $\dopp{R}$, but these are irreducible. Hence, $W$ is
  a unitary operator. 

  $Ad_W$ induces a unitary equivalence of the respective local
  algebras associated with every $I\Subset\dopp{R}$ by its definition
  (\ref{eq:defcospauni}) and the separating property of the
  vacuum. Furthermore, $W$ is readily shown to be covariant. If we
  denote the covariance automorphisms of
  $\lok{A}e_\lok{A}\otimes \lok{C}e_\lok{C}$ by $\alpha^\otimes$,
  we have for $gI\Subset\dopp{R}$, $I\Subset\dopp{R}$:
  $\alpha^\otimes_g Ad_W|_{\lok{A}\vee\lok{C}(I)}  = Ad_W 
  \alpha_g|_{\lok{A}\vee\lok{C}(I)}$. 
Using the \Name{Reeh-Schlieder}
  property of the local algebras, one may reconstruct the
  representations $Ue_\lok{A}\otimes Ue_\lok{C}(.)$ and
  $U(.)e_{\lok{A}\vee\lok{C}}$ from the action of the
  automorphisms.
  This, in turn, proves that $W$ intertwines the representations
  $U(.)e_{\lok{A}\vee\lok{C}}$ and $Ue_\lok{A}\otimes
  Ue_\lok{C}(.)$. Finally, we reconstruct the conformal models from
  their restrictions to the light-ray using conformal covariance.

In the following discussion $A$ denotes local observables in
$\lok{A}\subset\lok{B}$ and $\pi_0(A)$ its representative in the
vacuum representation on $e_\lok{A}\Hilb{H}=:\Hilb{H}_0$. The
implementation of conformal covariance in $\pi_0$ shall be written $U_0$. 
For every vacuum subrepresentation in
$\lok{A}\subset\lok{B}$ there is a partial isometry $R: \Hilb{H}
\rightarrow \Hilb{H}_0$ satisfying $RA = \pi_0(A)R$, for all local $A$ in
$\lok{A}\subset\lok{B}$.  

The projection $e_R:=R^*R$ commutes with all of
$\lok{A}$. $RU^\lok{A}(.)R^*$ is a unitary strongly continuous
representation of $PSL(2,\dopp{R})^\sim$ which implements global
conformal covariance in $\pi_0$, thus: $RU^\lok{A}(.)R^*= U_0(.)$. It
follows directly that $\Phi_\Omega:=R^*\Omega$, the vacuum of the
subrepresentation associated with $R$, is invariant with
respect to $U^\lok{A}$, ie $\Phi_\Omega\in E_\lok{A}\Hilb{H}$. This
completes the proof of the last statement.     
 \end{pf}

It is not clear in general that the representation
$\lok{A}\vee\lok{C}_{max}\subset\lok{B}$ of the tensor-product theory
defined by the vacuum  representation of a chiral subnet
$\lok{A}\subset\lok{B}$ and the vacuum representation of its maximal
\Name{Coset} model has a (spatial) tensor-product decomposition. This
is known under certain conditions \cite{KLM01}.   We write
$\lok{A}\otimes\lok{C}$ for the vacuum representation of
$\lok{A}\vee\lok{C}$. 

We  now give a characterisation of isotony for the local relative
commutants. The statement on the  cyclic projections is non-trivial
since, although the local relative commutants are manifestly covariant
with respect to $U$, the \Name{Reeh-Schlieder} theorem does not apply
due to the unclear status of isotony.  
\begin{prop}\label{prop:chariso}
  Assume  the unique inner-implementing representation $U^\lok{A}$
  associated with a chiral subnet 
  $\lok{A}\subset\lok{B}$ to have the net-endomorphism
  property. Referring to $I\Subset\Seins$, $e^c_I$ shall denote the projection
  onto the \Name{Hilbert} subspace which the local relative
  commutant $\lok{C}_I = \lok{A}(I)'\cap\lok{B}(I)$ generates from the
  vacuum. The     following are equivalent: 
  \begin{enumerate}
  \item \label{chisoi}For some pair $I,K$ of intervals satisfying
    $K\subsetneq I\Subset\Seins$ holds: $e^c_K\subset  e^c_I$. 
  \item \label{chisoii} $\lok{C}_{\Seins_+}   \subset
  \{U^\lok{A}(\tilde{D}(t)), \, t\in\dopp{R}\}'$. 
  \item \label{chisoiii} $\lok{C}_{max}(I) = \{U^\lok{A}(\tilde{g}),
  \, \tilde{g}\in\PSL(2,\dopp{R})^\sim\}'\cap
  \lok{B}(I) =  \lok{C}_I$, $I\Subset\Seins$.
  \end{enumerate} 
\end{prop}
Remark: The statement on the  cyclic projections is non-trivial
since, although the local relative commutants are manifestly covariant
with respect to $U$, the \Name{Reeh-Schlieder} theorem does not apply
due to the unclear status of isotony (cf eg \cite{hjB68}).

\begin{pf}
  The implications \ref{chisoiii} $\Rightarrow$  \ref{chisoi},
  \ref{chisoii} are obvious.  We start the proof proper with a
  discussion on \ref{chisoi} $\Rightarrow$ \ref{chisoiii}  and here we
  look at the case $I=\Seins_+$ (general case by covariance). We set
  $e^c_{\Seins_+}= e^c_+$. The inclusion $e^c_K\subset  e^c_+$ yields
  by the separating property of the vacuum and modular covariance of
  $\lok{C}_{\Seins_+}\subset \lok{B}(\Seins_+)$:  
  $\lok{C}_K\subset\lok{C}_{\Seins_+}$. Thus, any $g\in
  \PSL(2,\dopp{R})$ satisfying $g\Seins_+ =K$ leads to an operator
  $U(g)$  which leaves $e^c_+\Hilb{H}$ globally invariant. $g$ has the
  form $g=  S(n)T(s)D(t)$, $n,s\geq 0$. $g$ may be chosen such that  $t=0$.

By modular covariance $J$, the modular conjugation of
$\lok{B}(\Seins_+)$, and $e^c_+$ commute and, by covariance and the
\Name{Bisognano-Wichmann} property of $\lok{B}$,  $Ad_{JU(R(\pi))}$ induces
an automorphism of $\lok{C}_{\Seins_+}$, so $e^c_+$ commutes with
$U(R(\pi))$, too. The relations $JT(s)J=T(-s)$, $JS(n)J=S(-n)$ lead to
$U(S(-n))U(T(-s))e^c_+\Hilb{H}\subset e^c_+\Hilb{H}$. 
We assume $n,s>0$ and define 
\begin{displaymath}
  g(n,s) := S\klammer{-n \, \frac{ns+(1+ns)^2}{2+ns}} \,\,
  T\klammer{-s \, \frac{2+ns}{ns+(1+ns)^2}} \,\, \klammer{S(n)T(s)}^2
  \,\, . 
\end{displaymath}
Applying scale covariance we arrive at: $U(g(n,s))e^c_+\Hilb{H}\subset
e^c_+\Hilb{H}$. The group element $g(n,s)$ leaves the point
$1\in\Seins$ invariant and is not a pure scale transformation. 
This proves that
all special conformal 
transformations leave $e^+_c$ invariant. The same follows for the
translations because of $R(\pi)S(n)R(\pi)=T(-n)$, which proves
$\komm{U(g)}{e^c_+}=0$    for all $g\in \PSL(2,\dopp{R})$ recognising
that translations and special conformal transformations generate the
whole group. For  $n=0$
or $s=0$ the last part 
applies directly. This proves: $e^c_K=e^c_+$ for all
$K\Subset\Seins$. By modular covariance of the inclusions
$\lok{C}_K\subset \lok{B}(K)$ we have $\lok{C}_K =
\{e^c_K\}'\cap\lok{B}(K)$ and this yields isotony for the local
relative commutants. The remainder follows by maximality of
$\lok{C}_{max}$.

Finally we discuss the implication \ref{chisoii} $\Rightarrow$
\ref{chisoiii}. If $B\in \lok{B}(\Seins_+)$ commutes with
$U^\lok{A}(\tilde{D}(t))$, $t\in\dopp{R}$, then $B\Omega$ is invariant
under the action of all of 
$U^\lok{A}$ (lemma \ref{lem:invvec}). If $\tilde{g}$ is sufficiently
close to the identity, $Ad_{U^\lok{A}(\tilde{g})}(B)$ is a local
operator (proposition \ref{prop:netend}),
and the separating property of the vacuum proves that $B$ commutes
with all of $U^\lok{A}$. Thereby, we arrive at 
$\lok{C}_{\Seins_+}\subset \lok{C}_{max}(\Seins_+)$, provided the
assumption in \ref{chisoii} holds. The other inclusion is trivial.  
\end{pf}

If the dilatations $U^\lok{A}(\tilde{D}(t))$, $t\in\dopp{R}$, induce
automorphisms of $\lok{B}(\Seins_+)$, the last part of the proof
shows  $\lok{C}_{max}(\Seins_+)$ to  be the fixed-point subalgebra with
respect to this automorphism group. Covariance leads to a
corresponding identification of  every $\lok{C}_{max}(I)$,
$I\Subset\Seins$. This may be regarded as an alternative ``local''
characterisation of $\lok{C}_{max}$, but since the automorphism groups
are determined by global observables, namely non-trivial unitaries
from $U^\lok{A}$, this is not satisfactory.

 Only for the final step of our analysis we need to invoke the
  \Name{Additional Assumption} once again:
\begin{lemma}\label{lem:isoUA}
  Assume the {\sc Additional Assumption}  to hold. Then we have:\newline
  $U^\lok{A}(\tilde{D}(t))\in \lok{A}(\Seins_+)\vee
  \lok{A}(\Seins_-)$, $t\in\dopp{R}$, and $U^\lok{A}$ has the
  net-endomorphism property.   
\end{lemma}
\begin{pf}
%
%
According to the {\sc Additional Assumption}  and lemma
\ref{lem:diffD} there exist, for small, fixed $t$, diffeomorphisms
  $g_\delta$, $g_\varepsilon$ localised in arbitrarily small
  neighbourhoods of $+1\in\Seins$ and $-1\in\Seins$, respectively, and
  diffeomorphisms $g_+^{\tau_1,\tau_2}$, $g_-^{\tau_1,\tau_2}$ which
  are localised in $\Seins_+$ and $\Seins_-$, respectively, and  phases
$\varphi(\tau_1,\tau_2)$ such that for $\tau_{1,2}\in\dopp{R}_+$:  
\begin{eqnarray*}
  U^\lok{A}(\tilde{D}(t))&=& \varphi(\tau_1, \tau_2) \, 
  \Upsilon^\lok{A}(\symb{p}^{-1}(g_+^{\tau_1,\tau_2})) \, 
  \Upsilon^\lok{A}(\symb{p}^{-1}(g_-^{\tau_1,\tau_2}))\\
  && \cdot \, Ad_{U^\lok{A}(\tilde{D}(\tau_1))}
    (\Upsilon^\lok{A}(\symb{p}^{-1}(g_\varepsilon)))  \,   
  Ad_{U^\lok{A}(\tilde{D}(-\tau_2))}
    (\Upsilon^\lok{A}(\symb{p}^{-1}(g_\delta))) \,\, .  
\end{eqnarray*}

Following \Name{Roberts} \cite[corollary 2.5]{jR74a}, dilatation
invariance of the vacuum and the shrinking supports ensure that the last
two operators converge weakly to their vacuum expectation values in the
limit $\tau_{1,2}\rightarrow \infty$. We rewrite the equation above:
\begin{eqnarray}
&&
  Ad_{U^\lok{A}(\tilde{D}(\tau_1))}(\Upsilon^\lok{A}(\symb{p}^{-1}(g_\varepsilon)))  
  Ad_{U^\lok{A}(\tilde{D}(-\tau_2))}(\Upsilon^\lok{A}(\symb{p}^{-1}(g_\delta)))
  U^\lok{A}(\tilde{D}(t))^* \nonumber\\
\label{eq:weakscalconv}
&=&\overline{\varphi(\tau_1, \tau_2)}
\Upsilon^\lok{A}(\symb{p}^{-1}(g_+^{\tau_1,\tau_2}))^*
  \Upsilon^\lok{A}(\symb{p}^{-1}(g_-^{\tau_1,\tau_2}))^*  \,\, .
\end{eqnarray}
The operators to the right converge weakly by this equation in the
limit $\tau_1, \tau_2 \rightarrow \infty$. For small
$t$, $g_\varepsilon$ and $g_\delta$ may be chosen 
close to the identity, $\omega(.)$ is continuous and normalised, which
means that for  $g_\varepsilon, g_\delta \approx id$ we have
$\omega(\Upsilon^\lok{A}(\symb{p}^{-1}(g_\varepsilon)))\neq 0$, 
$\omega(\Upsilon^\lok{A}(\symb{p}^{-1}(g_\delta)))\neq 0$.  
This implies  $U^\lok{A}(\tilde{D}(t))\in \lok{A}(\Seins_+)\vee
  \lok{A}(\Seins_-)$ for small and hence for all $t$.

Because $\Upsilon^\lok{A}(\symb{p}^{-1}(g_+^{\tau_1,\tau_2}))$ and
$\Upsilon^\lok{A}(\symb{p}^{-1}(g_-^{\tau_1,\tau_2}))$ are unitary
operators, the right-hand side of equation (\ref{eq:weakscalconv})
converges, up to a phase, strongly against $U^\lok{A}(\tilde{D}(t))$
for small $t$. This strong convergence proves that for
$B\in\lok{B}(\Seins_+)$ and small $t$ holds true in the weak topology:
\begin{equation}
  \label{eq:denetend}
  U^\lok{A}(\tilde{D}(t)) B U^\lok{A}(\tilde{D}(t))^* =
  \lim_{\tau_1,\tau_2\rightarrow \infty}
  Ad_{\Upsilon^\lok{A}(\symb{p}^{-1}(g_+^{\tau_1,\tau_2}))} (B) \in
  \lok{B}(\Seins_+) \,\, .
\end{equation}
This establishes the net-endomorphism property (definition \ref{def:netend}).
\end{pf}

 The statement of this lemma holds trivially, if the global
algebra $\lok{A}$      coincides with
$\lok{A}(\Seins_+)\vee\lok{A}(\Seins_-)$. This is a desirable property
(eg for the {\em \Name{Connes}' fusion} approach to superselection
structure) and it holds in presence of strong additivity, but a
proof of it relying on general properties of chiral  conformal
subtheories seems out of reach.

We summarise and state the main result of this work, which
proves that the maximal \Name{Coset}  models are of a {\em local nature}:
\begin{satz}[main theorem]\label{th:main}
  $\lok{A}\subset\lok{B}$ an inclusion of chiral conformal quantum
  theories and suppose the  \Name{Additional Assumption} to hold. Then the
  unique  inner-implementing representation $U^\lok{A}$ has the
  net-endomorphism property and for all $I\Subset\Seins$ holds:
  \begin{displaymath}
    \lok{C}_{max} (I) := \{U^\lok{A}\}' \cap \lok{B}(I) = \lok{C}_I :=
    \lok{A}(I)' \cap \lok{B}(I)\,\,. 
  \end{displaymath}
\end{satz}

\begin{pf}
  The net-endomorphism property of $U^\lok{A}$ holds by lemma
  \ref{lem:autUA}, and \ref{chisoii} in proposition \ref{prop:chariso}
  is fulfilled  because of lemma \ref{lem:isoUA}.
\end{pf}

 In the cases where both $\lok{A}$ and $\lok{B}$ possess
  an integrable stress-energy tensor, and hence $\lok{C}_{max}$ alike,
  the main theorem means in particular: $\lok{A}_{max}(I)$ and
  $\lok{C}_{max}(I)$, $I\Subset\Seins$ arbitrary, are their mutual
  relative commutants in $\lok{B}(I)$.   The local algebras
  $\lok{A}_{max}(I)$ are factors which shows the relative commutant of 
  $\lok{A}_{max}(I)\vee\lok{C}_{max}(I)$ in $\lok{B}(I)$ is
  $\dopp{C}\Einsop$, ie this inclusion is irreducible.

The main theorem proves the conclusions of \Name{Rehren} \cite{khR00} to
hold true which rely on the {\em generating property} of nets of chiral
observables, if the $\opo$-dimensional theory contains a stress-energy
tensor in the sense of the \Name{L\"uscher-Mack} theorem
\cite{FST89}. Since such a stress-energy tensor factorises into its 
independent chiral components, our analysis applies directly. The
{\em generating property} introduced  in \cite{khR00} resisted 
attempts of proof even in presence of a stress-energy tensor,
unfortunately. 


Further remarks\footnote{Further details in \cite{sK03d}  and in the
  appendix.}: Results of \Name{Xu} \cite{fX00, fX00a}, \Name{Longo}
\cite{rL01} show that the current algebras $SU(n)_k$, $n,k\in
\dopp{N}$ and all
$\mathcal{V}ir_{c<1}$ models\footnote{This list may easily extended,
  eg by looking at branching rules as eg in \cite{KW88, KS88} and
  through conformal inclusions.}
are {\em completely rational} \cite{KLM01}, ie they 
have finitely many sectors,  all with finite statistics, they are 
strongly additive and satisfy the split property. Finiteness of statistics
 shows that the decomposition formulae of \Name{Kac and
  Wakimoto} of inclusions in the current algebras just mentioned yield
examples of nets of normal, irreducible {\em canonical tensor product
subfactors} (normal \Name{CTPS}) in the sense of \Name{Rehren} \cite{khR00}. 
For these inclusions the fact that $\lok{A}_{max}$ and $\lok{C}_{max}$
are locally their mutual relative commutants follows from the
heredity of strong additivity for inclusions of finite index
\cite{rL01}.  
Our  
result gives an independent proof relying on the presence of
stress-energy  tensors only and covers directly all current algebra
inclusions. 

\Name{Rehren} has shown for normal \Name{CTPS}  that the sectors
$\rho^{\lok{A}_{max}}_i\prec \rho$,
$\rho^{\lok{C}_{max}}_j\prec \rho$ form sets which are closed
under conjugation and (up to direct sums) fusion, that the coupling
matrix $Z_{ij}$ has to be a 
permutation matrix and
that the coupling matrix induces an isomorphism of the fusions rules
of $\lok{A}_{max}$ and $\lok{C}_{max}$ as far as only sub-endomorphisms
of $\rho$ are involved.  In particular, the statistical dimensions of
$\rho^{\lok{A}_{max}}_i$  and
$\rho^{\lok{C}_{max}}_{j}$ have to coincide for $Z_{ij}\neq 0
\Rightarrow Z_{ij}=1$. Thereby,
the results of \Name{Kac and Wakimoto} and similar decomposition
formulae allow us to translate information  on the superselection
structure  of $\lok{A}_{max}$ into  information on $\lok{C}_{max}$ and
vice versa.

Recently, \Name{M"uger} \cite{mM03} succeeded in extending the results
of \Name{Rehren}: He proved that for normal \Name{CTPS}
$\lok{A}_{max}\vee\lok{C}_ {max}\subset\lok{B}$  the coupling matrix
induces even an isomorphism of the respective \Name{DHR} subcategories,
if $\lok{B}$ has trivial superselection structure, ie the vacuum
representation  is its only locally normal representation. There is
one current algebra which has trivial superselection structure, namely
$E(8)_1$.  

In a study on branching rules associated with conformal inclusions in
exceptional current algebras, \Name{Kac and Niculescu Sanielevici} \cite{KS88}
provided some decomposition formulae which yield  examples of this
structure. Particularly interesting is the embedding $SU(2)_{16}\vee
SU(3)_6\subset E(8)_1$. If we regard $SU(2)_{16}$ as chiral conformal
subtheory $\lok{A}\subset E(8)_1$ and $ SU(3)_6$ as associated
\Name{Coset} model, $\lok{C}$, then both $\lok{A}_{max}$ and
$\lok{C}_{max}$ are non-trivial extensions and the localised
representation connected with $\lok{A}_{max}\vee\lok{C}_{max}\subset
E(8)_1$ is found to be  a ``diagonal'' sum of six tensor
products. The latter are known to be inequivalent by the result of
\Name{Rehren} and \Name{M\"{u}ger}'s results show, that the  the
respective \Name{DHR} categories associated with the 
endomorphisms of $\lok{A}_{max}$ and $\lok{C}_{max}$, respectively,
involved in $\rho$ are isomorphic.

\section{Discussion}
\label{sec:diss}

Information on the way in which the \Name{Borchers-Sugawara}
representation $U^\lok{A}$ associated with a chiral
conformal subtheory $\lok{A}\subset\lok{B}$ is generated by local
observables led  to further knowledge on $U^\lok{A}$. This, in turn,
was exploited for proving the maximal \Name{Coset} model,
$\lok{C}_{max}$, associated with $\lok{A}\subset\lok{B}$  to be of a
{\em local nature}, more specifically to coincide with the respective local
relative commutant. This way, we provided a solution of the {\em isotony
problem} for a large class of chiral subsystems
$\lok{A}\subset\lok{B}$ (theorem \ref{th:main}). 

All that turned out to  be necessary were two special features of the
implementers of dilatations 
(lemmas \ref{lem:autUA} and \ref{lem:isoUA}). 
The first one leads to an understanding how
$U^\lok{A}$ acts on general local observables in $\lok{B}$
geometrically which we summarised as {\em net-endomorphism property}
(proposition \ref{prop:netend}, definition \ref{def:netend}). We found
this property is in complete correspondence with the geometry of a
$\opo$-dimensional conformal quantum theory and derived all properties
one can ask from a $\opo$-dimensional holographic image of $\lok{B}$
(theorem \ref{th:chihol}). The derivation of the
net-endomorphism property relied mainly on a result on the interplay
of modular theory and positivity of energy; we found it worth while to
summarise and reformulate the facts known today in a natural converse
of \Name{Borchers}' theorem on half-sided translations (theorem
\ref{bowiesatz}). 

Our solution of the isotony problem made use of specific structures
of the chiral conformal group, $\PSL(2,\dopp{R})^\sim$, and
integrable positive energy representations of the group of orientation
preserving diffeomorphisms of the circle, $\Diff_+(\Seins)$. The
results are satisfactory in many respects: 

The \Name{Borchers-Sugawara} construction of $U^\lok{A}$ is completely
general, yet completely independent of local information and we have
shown that additional input is needed only for deriving two natural
lemmas (lemmas \ref{lem:autUA} and \ref{lem:isoUA}). 

Our  \Name{Additional Assumption}, the presence of an integrable
stress-energy tensor, is satisfied for a large class of well
investigated examples, 
the inclusions of current algebras (cf appendix). The main results
exhibit the natural objects of studies on \Name{Coset} models  to be
the maximal \Name{Coset} model, $\lok{C}_{max}$ and the maximal
covariant extension, $\lok{A}_{max}$, and opened the gate for a direct
incorporation of  results which have  been compiled in
research in representation theory of affine \Name{Lie} algebras and
string theory. In particular, we made accessible examples
of normal canonical tensor product subfactors \cite{khR00} in which
$\lok{A}_{max}$ and $\lok{C}_{max}$ are both non-trivial local extensions.

Yet, there are chiral conformal models which do not possess
a stress-energy tensor \cite{sK03d}, so our analysis asks for a more
general approach. The most general concept relating
covariance and local observables is given in terms of {\em local
implementers}  constructed via the universal localisation map as an
application of the split property, known as the quantum \Name{Noether}
theorem \cite{BDL86}. Especially in connection with chiral conformal
models this concept proved applicable: \Name{Carpi} \cite{sC99a}
reconstructed the stress-energy tensor of certain models via point
like limits of local implementers by methods which were introduced and
applied in the context of general chiral conformal quantum theory by
\Name{Fredenhagen and J\"or\ss} \cite{FJ96} with remarkable success. 

Approaching the problem from this angle appears to be
promising. First, we have reduced it to a question on the way a
particular set of global observables, namely to the dilatation group
$U^\lok{A}(\tilde{D}(.))$, is generated by local observables  of
$\lok{A}\subset \lok{B}$. The dilatations proved natural and  very
useful to look at 
in connection with the isotony problem. Secondly, the models to
look at first, the conformally covariant derivatives of the $U(1)$ current,
obey canonical commutation relations and are well known in many
respects (see eg \cite{jY94, GLW98}). Analytic problems
connected  with nuclearity and the split property have been addressed
successfully for free fields \cite{BW86,ADF87}, and the task looks
interesting and difficult   enough.

As we already mentioned, our analysis does not directly extend to
subsystems in other spacetimes. Conformally invariant theories in
higher dimensions might be accessible by the more general applicability
of the \Name{Borchers-Sugawara} construction \cite{sK02} and the
presence of spacetime symmetry groups leaving compact localisation
regions globally fixed (cf eg \cite{BGL93}), so an analysis based on
local implementers might work here as well. For other local quantum
theories the isotony problem has been solved by methods less direct
than ours, but very general ones \cite{CC01,CC03}. Thus, the quest
for the heart of this problem still awaits further investigation.

\subsection*{Acknowledgements}
I thank \Name{K.-H.\ Rehren} and \Name{S. Carpi} for
interesting discussions and helpful criticism.
 Financial support from the \Name{Ev.\ Studienwerk
    Villigst} is gratefully acknowledged.

\section*{Appendix}
\label{app}

The first part of this appendix discusses the  
background on our additional assumption. Since we only need to give a
summary of (mostly) well known results, the discussion  will be
brief; further details may be found in \cite{sK03d}. The second part contains a simple lemma on the position of
dilatations (scale transformations) in $\Diff_+(\Seins)^\sim$. 

Chiral current algebras provide a large class of interesting
models. They constitute chiral conformal quantum field theories
defined by fields, the currents, for which the commutator is linear in
the fields. The current algebras which we are interested in  are
labelled by reductive, compact \Name{Lie} algebras, which we call the
respective {\em colour algebra} of the model. The structure constants
of the colour algebra determine the current algebra up a central
extension, which is labelled for any simple ideal in the colour
algebra by a positive integer, called the {\em level}.

The current algebras associated with abelian and with simple, non-abelian
colour algebras of type $A(n), B(n), C(n), D(n), G(2)$ can be
constructed at level $1$ as 
{\em quark models} \cite{BH71}, ie by combining  \Name{Wick} squares of free
fermions  (reviews: \cite{GO86,FST89}). The corresponding models of
higher level, 
$k$, are constructed by tensoring the level $1$ current algebra $k$
times with itself and extracting the vacuum representation of the
level $k$ current (sub-)algebra from this tensor product
representation. We denote the current algebra of type $A(n)$ at level
$k$ as $A(n)_k$ and extend this notation to all other simple,
non-abelian current algebras mutatis mutandis. 

These models fulfill manifestly \Name{Wightman}'s axioms. The
conformal \Name{Hamilton}ian yields an energy grading on them in terms
of {\em modes}. In the mode  picture, the current algebra associated with
a simple, non-abelian colour algebra is given by commutation relations of
the following form:
\begin{displaymath}
  \komm{j^a_m}{j^b_n} = i f^{ab}{}_c j^c_{m+n} + k g^{ab} m
  \delta_{m,-n}\,\, .
\end{displaymath}
The indices $a,b,c$ refer to a basis of the colour \Name{Lie} algebra, 
$f^{ab}{}_c$ are its structure constants, $g^{ab}$ its \Name{Cartan}
metric (in a natural normalisation). Unitarity reads in terms of the
modes $j^a_n$: $j^a_n = j^a_{-n}{}^\dagger$. 

Using these commutation relations and positivity of energy it is
possible to establish linear $H$ bounds by arguments as in
\cite{BS90}. This yields the following properties of the 
fields: smeared currents which are symmetric and localised in a proper
interval are essentially self-adjoint on the \Name{Wightman} domain
and generate local \Name{v.Neumann} algebras which satisfy the
\Name{Haag-Kastler} axioms of local quantum theory
\cite{HK64, rH92}. For current algebras with an abelian colour algebra these
axioms were established by \Name{Buchholz and Schulz-Mirbach}
\cite{BS90}.

For the cases $E_6$, $E_7$,
 $E_8$, $F_4$ there are means to construct the current algebras in the
 mode picture at
 all levels, and  \Name{Wightman}'s axioms appear 
 implicitly in the literature for these cases (see \cite{vKbook}, \cite{GW84}, \cite{vL97}). Hence, the
 \Name{Haag-Kastler} axioms may be proved as indicated above. Another
 way of establishing current algebras as local quantum theories is to
 look at the exponentiated, positive-energy representations of loop
 groups stemming from the modes\footnote{This approach
  relies entirely on the structure and representation theory of loop
  groups and their \Name{Lie} algebras and works for the integration
  of quark models as well. Mentioning the integration
  through linear $H$ bounds seems worth while since this is closer to
  the general approach for the transition from a local quantum theory in terms
  of quantum fields to the corresponding formulation in terms of
  local algebras of bounded operators.} \cite{FG93,vL97,aW98}.

In conformally covariant \Name{Wightman} quantum field theories in one
(chiral) or $1+1$ dimensions the theorem of \Name{L\"uscher and Mack}
\cite{FST89} determines the commutation relations of a symmetric
\Name{Wightman} 
field implementing the infinitesimal conformal spacetime
transformations on fields of the quantum field theory, the {\em stress
  energy tensor} (\Name{SET}),  up to a numerical constant, which is
determined by 
the two-point function of the \Name{SET}. In $1+1$ dimensions the
\Name{SET} is found to factorise into two (independent)
chiral components. The chiral \Name{SET}s form, by their commutation
relations, an infinitesimal, positive energy representation of
$\Diff_+(\Seins)$. In terms of its modes, $L_n$, the \Name{SET} defines a
\Name{Virasoro} algebra with the  numerical constant, the central
charge $c$, determining 
the central extension:     
\begin{displaymath}
  \komm{L_m}{L_n} = (m-n) L_{m+n} + (m-1)m(m+1) \frac{c}{12}
  \delta_{m,-n} \,\, .
\end{displaymath}    
Unitarity of this representation manifests itself in the relations
$L_n = L_{-n}^\dagger$. Either by establishing linear $H$ bounds
\cite{BS90} or by integrating 
the \Name{Virasoro} algebra \cite{GW85,tL94} the \Name{SET} can
be shown to define a conformally covariant, local quantum
theory. 

In both formulations of current algebras the (\Name{Segal-})
\Name{Sugawara} construction (cf \cite{PS86} (\S 9.4), \cite{FST89}) yields a
\Name{SET} which is a quadratic function of the currents, either in
terms of the modes or as a \Name{Wick} square. In fact, these  models
prove to be diffeomorphism invariant. 

The embedding of one colour \Name{Lie} algebra, $\lie{h}$, into
another one, $\lie{g}$, yields an inclusion of current algebras and the
respective chiral conformal quantum theories. By the (\Name{Segal}-) 
\Name{Sugawara} construction both current algebras contain a
\Name{SET}. Due to complete reducibility results \cite{KW88,vKbook}
the respective 
current algebra associated with $\lie{h}\subset\lie{g}$ 
and the  \Name{Sugawara}-\Name{Virasoro} algebras are known to be
represented as direct sums of irreducible highest-weight
representations tensored by trivial representations on multiplicity spaces.

This ensures the integrability of the infinitesimal representations
and even more: the cocycles of the respective representations are
found to be completely determined by the infinitesimal central
extensions, ie the cocycles for all irreducible subrepresentations
agree and the group laws are fulfilled in the direct sum
representations up to {\em phases}. Due to technical problems
connected with the infinite dimension of the groups/ \Name{Lie}
algebras involved  this is not obvious at all, but it has been
established by \Name{Toledano-Laredo} \cite{vL99a} on the grounds
indicated here.  

This is of special importance for the \Name{SET} of the current
algebra associated with $\lie{h}$ embedded in the current algebra of
$\lie{g}$. It is straightforward to prove that the modes $L_{\pm 1,
  0}^{\lie{h}\subset\lie{g}}$ agree with the respective linear
combinations of the generators of $U^\lok{A}$, where $\lok{A}\subset
\lok{B}$ is taken to be the corresponding inclusion of current
algebras as local quantum theories. This identity  follows by
integrability of both infinitesimal representations (eg \cite{jF77}) and uniqueness
of $U^\lok{A}$ \cite{sK02}.

Our  additional assumption is thus shown to hold in this class of
examples: $\Upsilon^\lok{A}$, the integrated representation
of $\Diff_+(\Seins)^\sim$ generated by the \Name{Sugawara} \Name{SET} of
$\lok{A}$, is a projective, unitary, strongly continuous, of
positive energy and its restriction to $\PSL(2,\dopp{R})^\sim$ agrees with
$U^{\lok{A}}$. The statement of 
the additional assumption possibly covers a more general set of
models,  but due to technical difficulties connected with the infinite
dimension of $\Diff_+(\Seins)^\sim$ one has to be content with 
discussing integrable representations.

We come now to a simple, technical lemma on the position of scale
transformations in $\Diff_+(\Seins)^\sim$. If $I_1$ and $I_2$ are
neighbouring intervals, the {\em completed union} which consists of
$I_1\cup I_2$ and the common boundary point will be denoted
$I_1\overline{\cup} I_2$; the result is assumed to be a proper interval in
$\Seins$.

\begin{lemma}\label{lem:diffD}
For a fixed scale transformation $D(t)\neq id$, $t$ small, there exist
diffeomorphisms  
$g_\delta,\,\, g_\varepsilon\in \Diff_+(\Seins)$ which are localised
in arbitrarily small 
neighbourhoods of $+1$ and $-1$, respectively, and which agree with
$D(t)$ close to $+1$ and $-1$, respectively, such
that, by defining 
\begin{displaymath}
  g^{\tau_1}_\delta 
:= D(\tau_1)g_\delta D(\tau_1)^{-1}
\, , \,\,
  g^{\tau_2}_\varepsilon 
:= D(\tau_2)^{-1}g_\varepsilon D(\tau_2)\,\, ,
\end{displaymath}
 we have for all $\tau_{1,2}\in\dopp{R}_+$:
 \begin{equation}\label{eq:scalposdiff}
   D(t) = g_+^{\tau_1,\tau_2} 
g_-^{\tau_1,\tau_2}   
g^{\tau_1}_\delta     g^{\tau_2}_\varepsilon
\,\, .
 \end{equation}
Here, the diffeomorphisms $g_+^{\tau_1,\tau_2}$, $ g_-^{\tau_1,\tau_2}$
are uniquely specified by their being localised in the upper and lower
half circle, respectively. After a local identification of
$\Diff_+(\Seins)$ with a sheet of $\Diff_+(\Seins)^\sim$ containing
the identity, equation (\ref{eq:scalposdiff}) still holds
for the respective images in $\Diff_+(\Seins)^\sim$.
\end{lemma}

\begin{pf}
If $I_1$ and $I_2$ are
neighbouring intervals, the {\em completed union} which consists of
$I_1\cup I_2$ and the common boundary point will be denoted
$I_1\overline{\cup} I_2$.

Choose a set $\{I^0_\iota, \iota= +,-,\delta,\varepsilon\}$ 
of proper, disjoint 
intervals such that ${I^0_\pm}\subset \Seins_\pm$, $+1\in
I^0_\delta$, 
$-1\in I^0_\varepsilon$, the $I^0_\iota$ are separated by proper
intervals $I_{a}$,.., $I_d$  and a covering of $\Seins$ by
proper intervals $I^1_\iota$ is defined through:
\begin{displaymath}
  I^1_+ := I_a\overline{\cup}I^0_+\overline{\cup}I_b\,,\,\,
  I^1_- := I_c\overline{\cup}I^0_-\overline{\cup}I_d\,,\,\, 
I^1_\delta:=I_a\overline{\cup}I^0_\delta\overline{\cup}I_d\,,\,\,
I^1_\varepsilon:=I_c\overline{\cup}I^0_\varepsilon\overline{\cup}I_b\,\, .
\end{displaymath}

 For fixed $t$, one can choose these intervals such that  $D(t)$
 satisfies $D(t) I^0_\iota \Subset I^1_\iota$. Since $D(t) \Seins_\pm \subset
\Seins_\pm$, we may choose $g_\delta$,
$g_\varepsilon$ close to $id$ such that $g_\delta$ agrees with $D(t)$
 on $I^0_\delta$ 
and with $id$ on $I^1_\delta{}'$ and $g_\varepsilon$ agrees with $D(t)$
on $I^0_\varepsilon$ 
and with $id$ on $I^1_\varepsilon{}'$. Referring to this choice we
set:
\begin{displaymath}
  g_\pm\restriction{}{I^1_\pm}:= D(t) g_\delta^{-1} g_\varepsilon^{-1}\restriction{}{\Seins_\pm}
  \,\,,\quad
g_\pm\restriction{}{\overline{\Seins_\mp}}  
:= id\restriction{}{\overline{\Seins_\mp}}\,.   
\end{displaymath}
Then we have $D(t)=g_+g_-g_\delta g_\varepsilon$. We may now apply the
definitions in the lemma to this choice 
and recognise the results to satisfy equally well the assumptions of
the construction 
just given. 

For a neighbourhood of the identity the covering projection $\symb{p}:
\Diff_+(\Seins)^\sim \rightarrow \Diff_+(\Seins)$ is a
homeomorphism. If we apply $\symb{p}^{-1}$ to $D(t)$,
$g_\delta$, $g_\varepsilon$, $g_+$, $g_-$, we have
$\symb{p}^{-1}(D(t)) = \symb{p}^{-1}(g_+) \symb{p}^{-1}(g_-)
\symb{p}^{-1}(g_\delta)  \symb{p}^{-1}(g_\varepsilon)$. For small
$\tau_1$, $\tau_2$ the equality (\ref{eq:scalposdiff}) holds with the
corresponding replacements, and the same is true for all
$\tau_{1,2}\in\dopp{R}_+$ by continuity: 
denoting the covering
projection from $\dopp{R}$ onto $\Seins$ by $\symb{p}$, all the group elements
involved belong to the identity component of the subgroup
of $\Diff_+(\Seins)^\sim$ which  stabilises $\symb{p}^{-1}(+1)$ and
$\symb{p}^{-1}(-1)$, ie we never leave the first sheet of the covering.   
\end{pf}


\end{document}